\documentclass{jfm}
\usepackage{lineno,hyperref}
\usepackage{hyperref}
\usepackage{fancyhdr}
\usepackage{amsmath}
\usepackage{mathtools}
\usepackage{amsfonts}
\usepackage{amssymb}
\usepackage{textcomp}
\usepackage{booktabs}                      % tables
\usepackage{multicol}
\usepackage{multirow}
\usepackage{xspace}
\usepackage{nomencl}
\usepackage{caption,tabularx,booktabs}
\usepackage{caption}
\usepackage{float}
\usepackage[dvipsnames]{xcolor}
\usepackage{rotating}
\makenomenclature
\usepackage{tabularx}                      % tables with predefined width
\usepackage[square,sort,comma,numbers]{natbib}
\usepackage{adjustbox}
\usepackage[graphicx]{realboxes}
\usepackage[toc,page]{appendix}
\usepackage{cancel}
\usepackage{graphicx}
\usepackage{subcaption}
\usepackage{epstopdf}
\usepackage{color}

\let\OldS\S
\renewcommand{\S}{\OldS\xspace}

\newcommand{\revA}[1]{{\color{black}#1}}

\newcommand{\revB}[1]{{\color{black}#1}}

\newcommand{\revC}[1]{{\color{black}#1}}

\shorttitle{Wind-forced breaking waves at high wind speed}

\shortauthor{Scapin et al.}

\title{Momentum fluxes in wind-forced \\ breaking waves} % tentative title
\author{
  Nicolò Scapin\aff{1,2},
  Jiarong Wu\aff{1,3},
  J. Thomas Farrar\aff{4},
  Bertrand Chapron\aff{5},
  Stéphane Popinet\aff{6} and
  Luc Deike\aff{1,2}  
}
\affiliation
{
\aff{1}Department of Mechanical and Aerospace Engineering, Princeton University, Princeton, NJ 08544, USA,
\aff{2}High Meadows Environmental Institute, Princeton University, Princeton, NJ 08544, USA,
\aff{3}Courant Institute of Mathematical Sciences, New York University, US,
\aff{4}Woods Hole Oceanographic Institution (WHOI),
\aff{5}IFREMER, Univ. Brest, CNRS, IRD, Laboratoire d'Océanographie Physique et Spatiale (LOPS), France,
\aff{6}Institut Jean Le Rond d’Alembert, CNRS UMR 7190, Sorbonne Université, Paris 75005, France.
}
\begin{document}

\maketitle
\begin{abstract}
%
% Abstract (tentative abstract)
%
We investigate the momentum fluxes between a turbulent air boundary layer and a growing-breaking wave field by solving the air-water two-phase Navier-Stokes equations through direct numerical simulations (DNS). A fully-developed turbulent airflow drives the growth of a narrowbanded wave field, whose amplitude increases until reaching breaking conditions. The breaking events result in a loss of wave energy, transferred to the water column, followed by renewed growth under wind forcing. We revisit the momentum flux analysis in a high-wind speed regime, characterized by the ratio of the friction velocity to wave speed $u_\ast/c$ in the range $[0.3-0.9]$, through the lens of growing-breaking cycles. The total momentum flux across the interface is dominated by pressure, which increases with $u_\ast/c$ during growth and reduces sharply during breaking. Drag reduction during breaking is linked to airflow separation, a sudden acceleration of the flow, an upward shift of the mean streamwise velocity profile, and a reduction in Reynolds shear stress. We characterize the reduction of pressure stress and flow acceleration through an aerodynamic drag coefficient by splitting the analysis between growing and breaking stages, treating them as separate sub-processes. While drag increases with $u_\ast/c$ during growth, it drops during breaking. Averaging over both stages leads to a saturation of the drag coefficient at high $u_\ast/c$, comparable to what is observed at high wind speeds in laboratory and field conditions. Our analysis suggests this saturation is controlled by breaking dynamics.
\end{abstract}
%
% Intro
%
\section{Introduction}\label{sec:intro}
%
% general intro on the topic with some references
% 
Ocean waves modulate the exchanges of mass, momentum, and energy at the ocean-atmosphere interface. Waves continuously modify the ocean surface roughness and affect the momentum transfer between the atmospheric and oceanic boundary layers~\citep{sullivan2010dynamics}. When waves break, sea spray droplets are produced~\citep{veron2015ocean}, together with the generation of bubbles due to the air entrainment~\citep{deike2018gas,deike2022mass}, enhancing heat and mass transfer~\citep{veron2015ocean,deike2022mass}. Wind forcing and ocean waves mutually influence each other. On the air side, the irregular and time-varying surface wave topography alters the wind velocity, air temperature, and humidity in space and time. On the waterside, the wind drives wave formation, promotes growth and steepening up to breaking, and drives upper-ocean turbulence. \par 
Wind-induced wave breaking plays a crucial role in the interaction between the ocean and the atmosphere. Wave breaking limits wave steepness, influences the momentum exchange between the atmospheric boundary layer and the upper ocean, and locally alters wind velocity profiles and those of any other tracers~\citep{melville1985momentum,melville1996role}. Wave breaking is a highly dissipative process that controls the energy transfer from the wind to water currents and the transition to turbulence in the upper ocean~\citep{lamarre1991air,melville1996role,veron2001experiments}, and influences the wave-induced Langmuir turbulence~\citep{mcwilliams1997langmuir,melville1998laboratory,mcwilliams2016submesoscale}. \par
Understanding the momentum and the energy exchanges in wind-forced breaking waves remains an active area of research, especially in the high wind speed regime above wind speed of $20-25$ $m/s$ (evaluated at $\mathrm{10 m}$ height). Accurate evaluation of the momentum fluxes (or wind stress) between the wind and the wave field is necessary to properly represent the turbulent boundary layers in the lower atmosphere and in the upper ocean. Without waves, the momentum flux at the ocean-atmosphere interface would be solely due to viscous effects. Waves introduce a pressure component in the flux, whose significance in the stress partitioning increases as wind speed and local wave steepness increase~\citep{edson2013exchange}, so that open ocean measurements report an increase in the momentum flux measured at the top of the wave boundary layer (conventionally taken at a $10$-m height), compared to a flat wall~\citep{edson2013exchange,ayet2022dynamical}. \par
In coupled ocean-atmosphere numerical models, the momentum flux $\tau$ is related to the wind speed at a given height using the drag coefficient, $C_D$
\begin{equation}\label{eqn:intro_cd}
  \tau = \rho_au_\ast^2 = \rho_a C_D U_{10}^2\mathrm{,}
\end{equation}
where $u_\ast$ is the friction velocity, $\rho_a$ the air density, and the reference velocity $U_{10}$ taken at a conventional reference height of $10$-m \cite{edson2013exchange}. Assuming a logarithmic wind velocity profile~\citep{monin1954basic,janssen2004interaction} in eq.~\eqref{eqn:intro_cd}, the calculation of the momentum flux $\tau$ is reduced to the estimation of the drag coefficient $C_D$. \par 
Field observations have shown that $C_D$ is strongly dependent on the wind velocity $U_{10}$~\citep{edson2013exchange,ayet2022dynamical}. However, significant scatter in the data is often observed at a given wind speed due to difficulty in making momentum flux measurements in the open ocean and the role of other variables than wind speed (waves, current) in controlling the drag~\citep{ayet2022dynamical}. The relationship between $C_D$ and $U_{10}$ is well constrained when the wind and the waves are at equilibrium at moderate wind speed, i.e. $U_{10} \leq 7-15$ $m/s$), but the scatter is larger at low wind speed due to misalignment between wind and waves~\citep{ayet2022dynamical,manzella2024reduction}. For wind speeds exceeding $U_{10}>20$ $m/s$, the $C_D$-$U_{10}$ relationship exhibits a saturation in the field and laboratory experiments~\citep{sroka2021review}, with considerable scatter~\cite{sullivan2010dynamics,donelan2004limiting,sroka2021review,sraj2013bayesian,curcic2020revised}. The precise saturation value and the underlying physical principle remain uncertain~\citep{takagaki2016mechanism,komori2018laboratory,chen2007cblast} while being a critical parameter to understanding the intensification of tropical~\citep{sroka2021review} and extra-tropical cyclones~\citep{gentile2021impact,gentile2022sensitivity}. The main hypotheses that have been proposed to explain the saturation of $C_D$ are: sea spray generation at high wind speed \citep{bye2006drag,veron2015ocean}, and airflow separation and flattening wave crests, and a marked reduction in surface roughness~\citep{donelan2004limiting}. In this latter scenario, the role of wave breaking on flow separation and modulating pressure stress remains to be quantified in order to better constrain momentum flux formulation~\citep{kudryavtsev2014impact}. \par
%
%Laboratory experiments and numerical simulations have examined the momentum flux and the partition of the interfacial stress in terms of the pressure and viscous contributions at moderate wind speed~\citep{buckley2020surface,yousefi2020momentum,sullivan2000simulation,yang2010direct,iafrati2019effects}. \revC{Separately,~\citet{yang2018direct} studied the disturbance of the air boundary layer induced by breaking waves using direct numerical simulations while ~\citet{lu2024numerical} considered the associated the heat transfer modulation}. \par
%
\revC{
Several studies have investigated momentum flux and the partitioning of interfacial stress into pressure and viscous contributions at moderate wind speeds, both experimentally, e.g.~\citep{buckley2020surface, yousefi2020momentum}, and numerically for idealized waves using three-dimensional Large Eddy Simulations e.g.~\citep{sullivan2000simulation,yang2010direct}. \citet{iafrati2019effects} and \citet{wu2021wind} considered two-phase Navier-Stokes equations in a two-dimensional configuration with idealized airflow, noting that the two-dimensional configuration can not capture the realistic turbulent boundary layer (lin-log velocity profile and three-dimensional fluctuations). ~\citet{yang2018direct} and~\citet{lu2024numerical} employed three-dimensional direct numerical simulations (DNS) to study the disturbance of the air boundary layer induced by breaking waves and to explore the associated modulation of heat transfer, while~\citet{wu2022revisiting} employed three-dimensional DNS to investigate momentum fluxes and wave growth for non-breaking wind waves.
} \par
So far, no fully-resolved simulations have been conducted for strongly forced wind wave breaking up to $u_\ast/c\approx 1$, including both the wind-wave growth stage up to breaking and the following breaking event. Such configurations, where breaking conditions are reached through wind forcing, would be ideal for studying the role of the breaking and growing wave stages on the turbulent airflow and momentum flux, permitting the separation of the contribution of both phenomena, which are usually time and ensemble-averaged in field and laboratory conditions. \par
Here, we consider the case of waves forced by a turbulent wind up to $u_\ast/c = 0.9$, including breaking events. The turbulent airflow and the water waves are fully resolved without any model for the wave shape, which can grow and break, and without any sub-grid turbulence model. This is accomplished by using a two-phase Navier-Stokes solver, with a geometric Volume-of-Fluid method to reconstruct the wave interface. For different values of $u_\ast/c$, we estimate the momentum fluxes by analysing the growing, breaking, and post-breaking stages of the wave field separately. Using this novel approach, we clarify the relation between the stress partition at the interface and the modulation of the velocity profile in the airflow region during the different stages of the wave dynamics. We extract an equivalent of the drag coefficient during the wave growth and breaking stages and discuss how their average values reproduce trends observed in the laboratory and field observations. \par
%
% Paper structure
%
This paper is organized as follows. In \S~\ref{sec:meth}, we introduce the overall methodology, including (i) the governing equations that we solve and a brief summary of the numerical algorithm (\S~\ref{sec:gov_eqn}), (ii) the numerical set-up (\S~\ref{sec:case_init}) and (iii) the physical dimensionless parameters, (\S~\ref{sec:num_setup}). In \S~\ref{sec:flow_visu}, we provide a qualitative description of the evolution of the coupled system composed of the wind and the growing/breaking wave field. In \S~\ref{sec:mom_flux}, we discuss the momentum flux over breaking waves and how the wave growth, steepening, and eventual breaking modulate the velocity profile and the Reynolds stress. In \S~\ref{sec:drag_coeff}, we connect our analysis to formulations of the drag coefficient over growing and breaking waves and compare them to experimental results. Conclusions are presented in \S~\ref{sec:concl}. 
%
% Methodology
%
\section{Methodology}\label{sec:meth}
In this section, we discuss the governing equations, numerical setup, and dimensionless parameters used to characterize the wind-forced wave growth and breaking processes. 
% Governing equation, case description, and simulation parameters
%
\subsection{Governing equations and numerical model}\label{sec:gov_eqn}
We investigate wind-forced breaking waves as a two-phase problem. We solve the incompressible, two-phase Navier-Stokes equations with surface tension, implemented in the open-source Basilisk solver\footnote{http://basilisk.fr/}~\citep{popinet2009accurate,popinet2015quadtree}, following the approach developed in~\cite{wu2021wind} and~\cite{wu2022revisiting} to study wind-wave growth. Briefly, to distinguish the two phases, an indicator function $\mathcal{F}$ is introduced and set equal to $1$ in the water phase within the volume $\Omega_w$ and $0$ in the air phase within the volume $\Omega_a$. The two domains are separated by a zero-thickness interface $\Gamma$. The indicator function $\mathcal{F}$ is governed by the transport equation~\citep{tryggvason2011direct}:
\begin{equation}\label{eqn:ind_fun}
  \dfrac{\partial\mathcal{F}}{\partial t} + \mathbf{u}\cdot\nabla\mathcal{F} = 0\mathcal{,}
\end{equation} 
where $\mathbf{u}=(u,v,w)$ is the one-fluid velocity, assumed continuous in the whole domain $\Omega=\Omega_w\cup\Omega_a$. The indicator function is employed to define the one-fluid density $\rho=\rho_w\mathcal{F}+\rho_a(1-\mathcal{F})$ and the one-fluid dynamic viscosity $\mu=\mu_w\mathcal{F}+\mu_a(1-\mathcal{F})$, where $\rho_w$, $\rho_a$ and $\mu_w$, $\mu_a$ are the densities and the dynamic viscosities of water and air, respectively. The transport of $\mathcal{F}$ is coupled with the incompressibility constraint and the Navier-Stokes equations for a Newtonian fluid. These equations, expressed in the one-fluid formulation~\citep{tryggvason2011direct}, read as follows
\begin{equation}\label{eqn:div_con}
  \nabla\cdot\mathbf{u} = 0\mathcal{,}
\end{equation}
\begin{equation}\label{eqn:mom_con}
  \dfrac{\partial(\rho\mathbf{u})}{\partial t} + \nabla\cdot(\rho\mathbf{u}\mathbf{u}) = - \nabla p +\nabla\cdot(2\mu\mathbf{D}) + \rho|\mathbf{g}|\mathbf{e}_z + \sigma\varkappa\delta_\Gamma\mathbf{n}\mathrm{,}
\end{equation}
where $p$ is the \revC{pressure}, \revB{$\mathbf{D}$ is the strain rate tensor}, i.e. $\mathbf{D}=\left(\nabla\mathbf{u}+\nabla\mathbf{u}^T\right)/2$, $|\mathbf{g}|$ is the modulus of the gravitational acceleration, $\mathbf{e}_z=(0,0,-1)$ is unit vector oriented like gravity, $\sigma$ is the surface tension coefficient, $\delta_\Gamma$ is a Dirac distribution function satisfying the identity $\int_\Gamma \delta_\Gamma dS=1$, $\mathbf{n}$ is the interface normal vector pointing outward to the liquid domain, and $\varkappa=\nabla\cdot\mathbf{n}$ is the interfacial curvature. \par 
Equations~\eqref{eqn:ind_fun},~\eqref{eqn:div_con} and~\eqref{eqn:mom_con} are solved using an adaptive mesh refinement (AMR) strategy on an octree grid, as implemented in Basilisk and described in~\citep{popinet2015quadtree,van2018towards}. The use of AMR significantly reduces the computational cost while efficiently representing different multiscale processes \citep{mostert2022high}. We employ a conservative and diffusion-free geometric volume-of-fluid (VOF) method~\citep{tryggvason2011direct} to discretize equation~\eqref{eqn:ind_fun}. The equation for the indicator function~\eqref{eqn:ind_fun} is then solved together with the momentum equation~\eqref{eqn:mom_con} using the Bell-Colella-Glaz method~\citep{bell1989second} for the advection part and a standard second-order finite-volume scheme for the viscous term. The viscous term is advanced semi-implicitly in time using a Crank–Nicolson scheme, whereas an explicit treatment is kept for the advection part. \par 
The momentum equation is discretized in a conservative and consistent manner~\citep{popinet2018numerical,mostert2022high} to ensure accurate and stable integration of equation~\eqref{eqn:mom_con} in the presence of a large difference in the density and viscosity of the two phases. The capillary and gravity forces are discretized using a well-balanced formulation~\citep{popinet2018numerical}, maintaining an exact equilibrium among pressure gradient, capillary, and gravity forces under static conditions, and minimizing the generation of artificial parasitic currents. These features are important for studying phenomena such as wave breaking~\citep{deike2015capillary,deike2016air,mostert2022high}, wind wave growth~\citep{wu2021wind,wu2022revisiting}, and, other classes of two-phase turbulent flows (see e.g.~\cite{riviere2021sub,perrard2021bubble,farsoiya2023role}).
\subsection{Configuration and initialization}\label{sec:case_init}
\begin{figure*}%[t!]
  \centering
  \includegraphics[trim={0cm 0.50cm 0cm 1.25cm}, clip, width=0.60\textwidth]{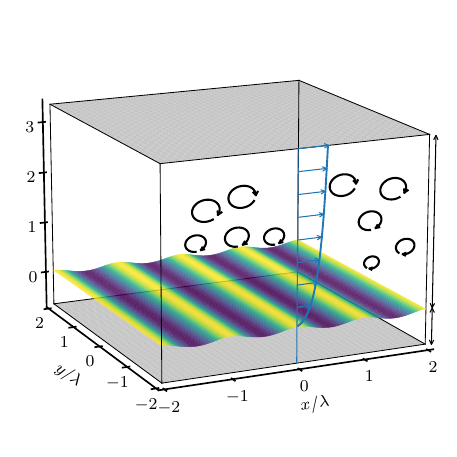}
  \put(-237,155.0){$z/\lambda$}
  \put(-005,050.5){$h_W/\lambda$}
  \put(-005,100.5){$(L_0-h_W)/\lambda$}
  \caption{Computational domain and physical configuration illustrating the initial condition for the airflow, wave and water field. The air and water mean heights are $(L_0-h_W)/\lambda$ and $h_W/\lambda$. The airflow is a fully developed turbulent boundary layer (mean profile in light-blue line, while turbulent eddies are illustrated in black). The wave field and the water region are initialized using an irrotational third-order Stokes wave solution~\citep{deike2015capillary}. The initial wave profile $\eta_0$ has zero spatial mean and a steepness $a_0k=0.3$. In the surface contour, dark-blue regions denote wave troughs, while yellow regions indicate wave crests.}
  \label{fig:sketch}
\end{figure*}

The governing equations~\eqref{eqn:ind_fun},~\eqref{eqn:div_con} and~\eqref{eqn:mom_con} are solved in a cubic computational box, as illustrated in figure~\ref{fig:sketch}, of size $L_0=4\lambda$, i.e. $\left[-2\lambda,2\lambda\right]\times\left[-2\lambda,2\lambda\right]\times\left[-h_W,4\lambda-h_W\right]$, where $\lambda$ is the wavelength and $h_W$ is the mean water depth. The streamwise, spanwise, and vertical directions correspond to the $x$, $y$, and $z$ directions, respectively. Periodic boundary conditions are prescribed for all the variables, i.e. $\mathbf{u}$, $p$, and $\mathcal{F}$, along the $x-y$ horizontal directions. On the top and bottom domain boundaries, which correspond to the grey geometric planes at $z=(L_0-h_W)/\lambda$ and $z=-h_W/\lambda$ in figure~\ref{fig:sketch}, a homogeneous Neumann boundary condition is prescribed for the pressure $p$, the indicator function $\mathcal{F}$ and the two horizontal velocity components $u$ and $v$, while a no-penetration boundary condition is imposed on the vertical velocity component $w$. \par
Figure~\ref{fig:sketch} illustrates the three domains of interest: the wave field occupies the region $z=\eta(x,y)$ and separates the water and air flows. The water flow occupies the region where $z<\eta(x,y)$, with a mean depth of $h_W$. The airflow occupies the region where $z>\eta(x,y)$ and has mean height $(L_0-h_W)$. The wave field, water, and air velocity must be properly initialized at the beginning of the simulation. \par
We initialize the wave and the associated water velocity field using a non-linear potential flow solution $\Phi_0$ with free surface $\eta_0$~\citep{lamb1993hydrodynamics}, and consider a third-order expansion of the system (third-order Stokes wave), with a given amplitude $a_0$ and wavenumber $k=2\pi/\lambda$. Once $\Phi_0$ is known, the initial velocity in the water $\mathbf{u}_{w,0}$ can be readily evaluated from the velocity potential as $\mathbf{u}_{w,0}=\nabla\Phi_0$, for the expressions of $\eta_0$ and $\Phi_0$ see~\citet{lamb1993hydrodynamics,deike2015capillary}. We note that this initialization prescribes an initial orbital velocity field in the water characterized by a zero Eulerian mean. It is worth mentioning that several previous works~\citep{chen1999two,iafrati2009numerical,deike2015capillary,deike2016air,yang2018direct,mostert2022high} have demonstrated the relevance of the Stokes waves solution for the simulation of breaking waves with numerical results that can be accurately compared to laboratory experiments. However, unlike the Stokes wave solution, which disregards surface tension effects, the current approach incorporates them. Thus, based on the dispersion relation, the wave speed reads as $c=\sqrt{|\mathbf{g}|/k + \sigma k/\rho_w}$. \revC{We consider the linear phase speed for gravity-capillary waves, as the typical nonlinear Stokes wave correction remains negligible during the initial growth stage. During the breaking stage, the wave speed undergoes a slight slowdown, which is not accounted for by this correction~\citep{banner2014linking}. However, the nominal wave speed set by the initial conditions remains an excellent approximation of the effective wave speed throughout the simulation.} \par
We initialize the airflow region using a fully developed turbulent flow field at the desired friction Reynolds number, as defined in section~\ref{sec:num_setup}, following \citet{wu2022revisiting}. This initialization involves a precursor simulation conducted independently in a single-phase setup and employing the same domain. During this precursor simulation, the wave field with profile $\eta_0$ remains at rest, and a no-slip/no-penetration boundary condition is enforced on the wave surface for the velocity field, using the embedded boundary method~\citep{johansen1998cartesian} available within the Basilisk framework~\citep{ghigo2021conservative}. The precursor simulation is performed long enough until the turbulent airflow achieves a statistically steady state by adding an external body force per unit mass acting in the streamwise direction, i.e. $\partial p_0/\partial x(1-\mathcal{F})\mathbf{e}_x$ with $\mathbf{e}_x=(1,0,0)$, on the right-hand side of the momentum equation~\eqref{eqn:mom_con}. In the expression of the body force, $\partial p_0/\partial x$ is a uniform pressure gradient driving the flow and, here, reads as
\begin{equation}\label{eqn:pif}
  \dfrac{\partial p_0}{\partial x} = \dfrac{\rho_au_\ast^2}{L_0-h_W}\mathrm{.}
\end{equation}
The imposed pressure gradient sets the nominal friction velocity $u_\ast$ and prescribes the total stress $\rho_au_\ast^2$ on the wave field. Once a statistically steady state is achieved for the precursor, the resulting fully developed turbulent field is employed as an initial condition for the airflow region in the two-phase simulations. \par 
Once the simulation begins, the airflow and wave field dynamically evolve without any prescribed conditions at the two-phase interface. This fully coupled system undergoes a short self-adjustment stage due to changes in boundary conditions at the interface and the motion of the water layer~\citep{wu2022revisiting}. As discussed by~\cite{wu2022revisiting}, the self-adjustment period is short compared to other processes of interest. In this work, we verify that the self-adjustment stage lasts less than $2\omega t$ in all cases, \revB{where $\omega=2\pi/T_0$ is the angular frequency and $T_0=\lambda/c$ is the wave period}; therefore, all analyses are conducted for $\omega t > 2$.
\subsection{Non-dimensional parameters}\label{sec:num_setup}
The flow field in the air and water phases depends on several dimensional parameters. Here, the number of independent physical dimensions is three, i.e. mass, length, and time. According to Buckingham's $\pi$ theorem, the $11$ physical variables ($\rho_a,\rho_w,\mu_a,\mu_w,L_0-h_W,h_W,\lambda,a_0,\sigma,|\mathbf{g}|,u_\ast$) can be reduced to $8$ dimensionless groups so that the problem can be described by the following dimensionless groups (note that other combinations would have been possible)
\begin{equation}\label{eqn:dim_group}
  \left(Re_{\ast,\lambda},Re_W,Bo,\dfrac{\rho_w}{\rho_a},\dfrac{L_0-h_W}{\lambda},\dfrac{h_W}{\lambda},a_0k,\dfrac{u_\ast}{c}\right)\mathrm{.}
\end{equation}
In~\eqref{eqn:dim_group}, $Re_{\ast,\lambda}=\rho_au_\ast\lambda/\mu_a$ and $Re_W=\rho_wc\lambda/\mu_w$ represent the friction and wave Reynolds numbers, respectively, reflecting the balance between inertia and viscous effects in the regions adjacent to the wave field, for the air and water phases. A third dependent friction Reynolds number is the one based on the size of air boundary layer $L_0-h_W$, i.e. $Re_{\ast}=Re_{\ast,\lambda}(L_0-h_W)/\lambda=\rho_au_\ast(L_0-h_W)/\mu_a$, which reflects the importance of inertia over viscous forces in the region well above the wave field. In Appendix~\ref{app:num_wave}, we show that $Re_{\ast,\lambda}$ is the physically relevant group in order to compare cases at different Reynolds numbers and, therefore, $Re_{\ast,\lambda}$ will be used in conjunction with $Re_W$. The Bond number $Bo=(\rho_w-\rho_a)|\mathbf{g}|/(\sigma k^2)$ provides the ratio between gravitational to restoring capillary forces and $\rho_w/\rho_a$ represents the density ratio, which is set to the one of water and air. Next, we have a set of ratios of length scale, or geometric parameters: $(L_0-h_W)/\lambda$ represents the ratio between the mean airflow height and the wavelength, $h_W/\lambda$ is the ratio between the mean height of the water depth and the wavelength $\lambda$. Sensitivity to these parameters is shown in the appendix~\ref{app:num_wave}, and we verify that the physical conclusions do not depend on their specific values. Finally, $a_0k$ represents the wave's initial steepness, and $u_\ast/c$ defines the ratio between the friction velocity and wave phase speed. \par
Here, we focus on the interaction between turbulent wind and breaking waves at high wind speed and systematically vary the ratio $u_\ast/c$ from 0.3 to 0.9 while also performing sensitivity tests in terms of $Re_{\ast,\lambda}$ and geometrical parameters, as discussed in Section~\ref{sec:mom_flux}. The range of $u_\ast/c$ from $0.3$ to $0.9$, is relevant to discuss small-scale physics of wind-wave fields during tropical cyclone conditions~\citep{sroka2021review}. \par
We consider four water waves and prescribe the airflow's mean height $(L_0-h_W)/\lambda = 3.36$. This value ensures that the air boundary layer is more than three times larger than one single wavelength, avoiding any confinement effect due to the top boundary, as discussed in ~\citet{wu2022revisiting}. We set the mean water depth $h_W/\lambda = 0.64$, which is considered adequate to satisfy the deep water assumption~\citep{deike2015capillary,deike2016air,yang2018direct}. We consider $Re_W=2.55\cdot 10^4$ and $Bo=200$, so that the waves are within the gravity regime. We consider the air-water density ratio, $\rho_w/\rho_a=816$, while the dynamic viscosity ratio (fixed through the choices of the Reynolds numbers) remains well above unity, as reported in the table~\ref{tab:set_up}. The initial wave steepness is $a_0k=0.3$, below breaking threshold, typically $[0.32-0.33]$~\citep{deike2015capillary} for the present configuration. This choice ensures that the wind drives the wave field to break. \par
In the present configuration, and unlike in laboratory experiments or open ocean conditions, $u_\ast/c$ is independent of $a_0k$, \revC{which allows us to isolate the effect of $u_\ast/c$ on the flow at the same $Re_{\ast,\lambda}$ (by varying the wave speed $c$ while keeping $u_\ast$ fixed from the same precursor).} 
Furthermore, we work at constant $Re_W$ so that the growth rate of the wave field is controlled by the wind forcing $u_\ast/c$, while the viscous dissipation is the same ~\citep{wu2022revisiting}, but note that the Reynolds number (air and water side) and Bond number we consider are significantly smaller than ocean waves with wavelength $1$ $m$ or larger due to computational limitations. The underlying assumption to compare our results to laboratory or field conditions is that we perform simulations in quasi-asymptotic regimes of high values of these parameters \citep{deike2016air,mostert2022high,wu2022revisiting}. The simulations are summarized in Table~\ref{tab:set_up}. \par
\begin{table}%[h!]
\centering
\begin{tabular}{ccccccc}
%\toprule
\toprule
{{$u_\ast/c$}} & {{$Re_{\ast,\lambda}$}} & {{$Re_W$}} & {{$\mu_w/\mu_a$}} & {{$(L_0-h_W)/\lambda$}} & {{$Bo$}} & {{$\mathrm{Le}$}} \\
\toprule
0.30 & 214  & $2.55 \cdot 10^{4}$ & 22.84 & 3.36 & 200 & 10 \\
0.40 & 214  & $2.55 \cdot 10^{4}$ & 17.13 & 3.36 & 200 & 10 \\
0.50 & 214  & $2.55 \cdot 10^{4}$ & 13.71 & 3.36 & 200 & 10 \\
0.70 & 214  & $2.55 \cdot 10^{4}$ & 9.79 & 3.36 & 200 & 10 \\
0.90 & 214  & $2.55 \cdot 10^{4}$ & 7.61 & 3.36 & 200 & 10 \\
\midrule
0.90 & 53.5 & $2.55 \cdot 10^{4}$ & 1.90 & 3.36 & 200 & 10 \\
0.90 & 107  & $2.55 \cdot 10^{4}$ & 3.81 & 3.36 & 200 & 10 \\
\midrule
0.90 & 214  & $2.55 \cdot 10^{4}$ & 7.61 & 3.36 & 200 & 11 \\
0.90 & 107  & $2.55 \cdot 10^{4}$ & 3.81 & 6.72 & 200 & 11 \\
%\bottomrule
\bottomrule
\end{tabular}
\caption{Summary of the simulated cases for different values of $u_\ast/c$. In the table, $Re_{\ast,\lambda}=\rho_au_\ast\lambda/\mu_a$, $Re_W=\rho_wc\lambda/\mu_w$, $\mu_w/\mu_a$, $(L_0-h_W)/\lambda$ and $Bo=(\rho_w-\rho_a)|\mathbf{g}|/(\sigma k^2)$ as defined in~\eqref{eqn:dim_group}. The cases with $(L_0-h_W)/\lambda=3.36$ corresponds to $4$ waves per box size, whereas the case with $(L_0-h_W)/\lambda=6.72$ to $8$ waves per box size. For the different cases, the initial steepness is set equal to $a_0k=0.3$, and the density ratio is taken as $\rho_w/\rho_a=816$.}
\label{tab:set_up}
\end{table}
%
%\par
%
The numerical grid in our simulations is adaptive, featuring a minimum grid size $\Delta = L_0/(2^\mathrm{Le})$, where $\mathrm{Le}$ represents the maximum level of refinement. The Adaptive Mesh Refinement (AMR) technique significantly reduces computational costs by maintaining a highly refined grid near the interface and in the boundary layers while allowing coarser grids in the bulk airflow, provided that refinement criteria are met~\citep{popinet2015quadtree,van2018towards}. The AMR technique has shown to be accurate for homogeneous and isotropic turbulent flow \citep{riviere2021sub,farsoiya2023role}, and wall-bounded turbulent flows and the present configuration in~\citep{wu2022revisiting}, where the grid is dynamically adapted with respect to the norm of the second-derivative of the velocity (in the air and the water phase) and of the volume fraction. Following ~\citet{wu2022revisiting}, the refinement criteria for the air and water velocity components and the volume fraction are set equal to $\epsilon_{ua}=0.3u_\ast$, $\epsilon_{uw} = 10^{-3}c$ and $\epsilon_\mathcal{F}=10^{-4}$, respectively. \par
In most simulations, $\mathrm{Le}=10$, with the maximum number of grid points reaching up to $70\cdot 10^6$, which corresponds to $7$ $\%$ of an equivalent $1024^3$ uniform grid. Additionally, we present in Appendix~\ref{app:conv} a grid refinement study up to $\mathrm{Le}=11$ for $u_\ast/c=0.9$. In this case, the maximum number of grid points is around $500\cdot 10^6$, which corresponds to about $6$ $\%$ of an equivalent $2048^3$ uniform grid. The significant reduction in the number of grid cells with AMR at $\mathrm{Le}=10-11$, compared to a uniform Cartesian grid, makes AMR an attractive approach for investigating wind-forced breaking waves. \par
The computational costs of these simulations include generating a precursor simulation and running the two-phase simulations. Four precursors are used for varying air-side Reynolds number ($Re_{\ast,\lambda}=53.5-107-214$) and the ratio $(L_0-h_W)/\lambda=3.36-6.72$, amounting to $\approx 1.5\cdot 10^5$ CPU each. Each two-phase simulation at $\mathrm{Le}=10$ requires $\approx 4.2\cdot 10^5$ CPU hours, while the two simulations at $\mathrm{Le}=11$ require $\approx 1.20\cdot 10^6$ CPU hours each. We employ $384$ processors for the precursor simulations and for the two-phase cases $480$ at $\mathrm{Le}=10$, and $980$ processors at $\mathrm{Le}=11$. The total cost of the simulation campaign is about $6\cdot 10^6$ CPU hours.
%
% Qualitative behaviour
%
\section{Evolution of wind-forced breaking waves}\label{sec:flow_visu}
This section discusses the evolution of the fully coupled system composed of the turbulent wind, the wave field, and the water column. 

\subsection{Wave interface evolution}
\begin{figure*}%[t!]
  \centering
  \includegraphics[trim={0cm 3.5cm 0cm 3.5cm}, clip, width=0.455\textwidth]{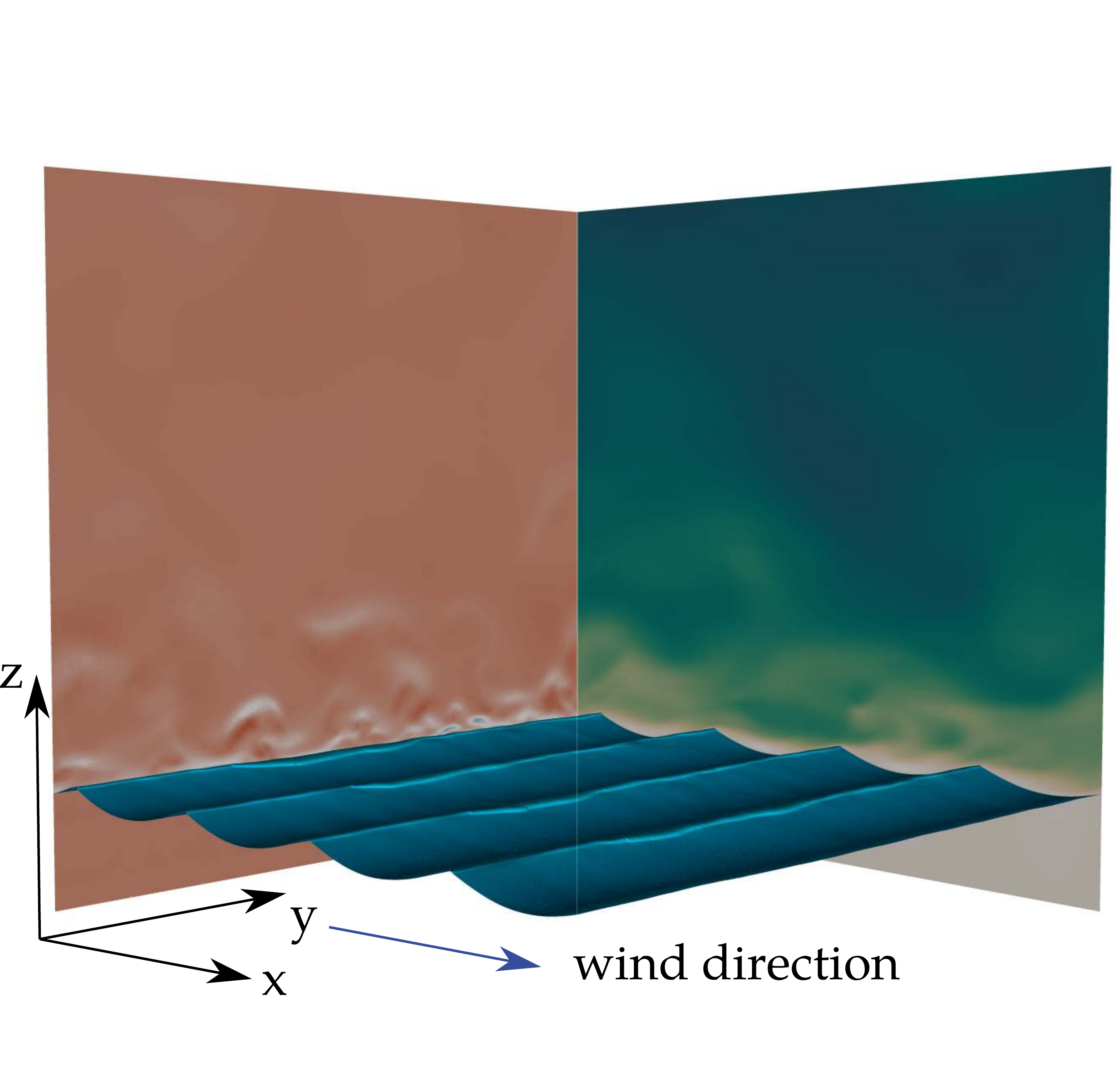}
  \hspace{0.25 cm}
  \includegraphics[trim={0cm 4.0cm 0cm 4.0cm}, clip, width=0.445\textwidth]{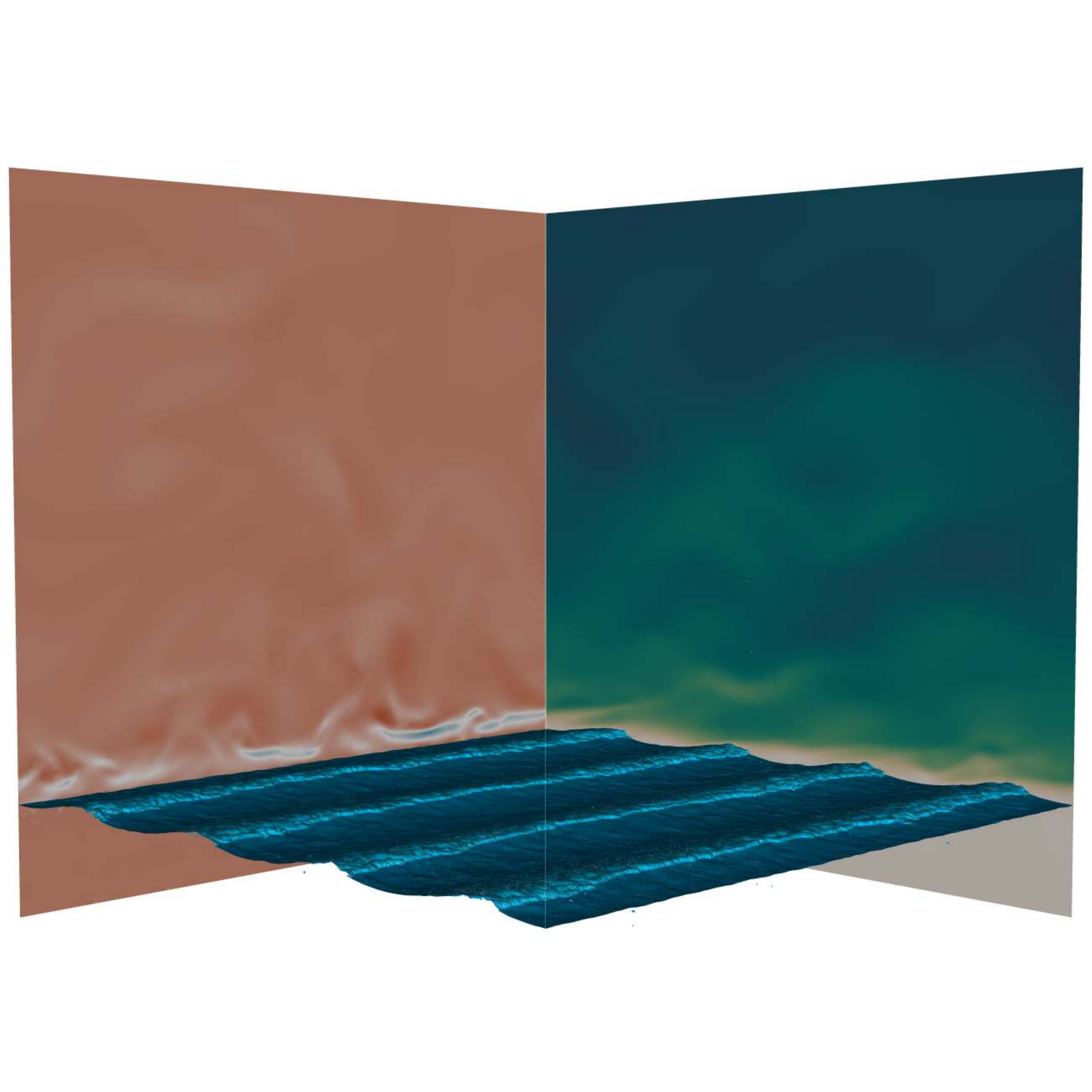}
  \put(-368,120){\small(a)}
  \put(-182,120){\small(b)} \\
  \vspace{0.5 cm}
  \includegraphics[trim={0cm 4.0cm 0cm 6.0cm}, clip, width=0.445\textwidth]{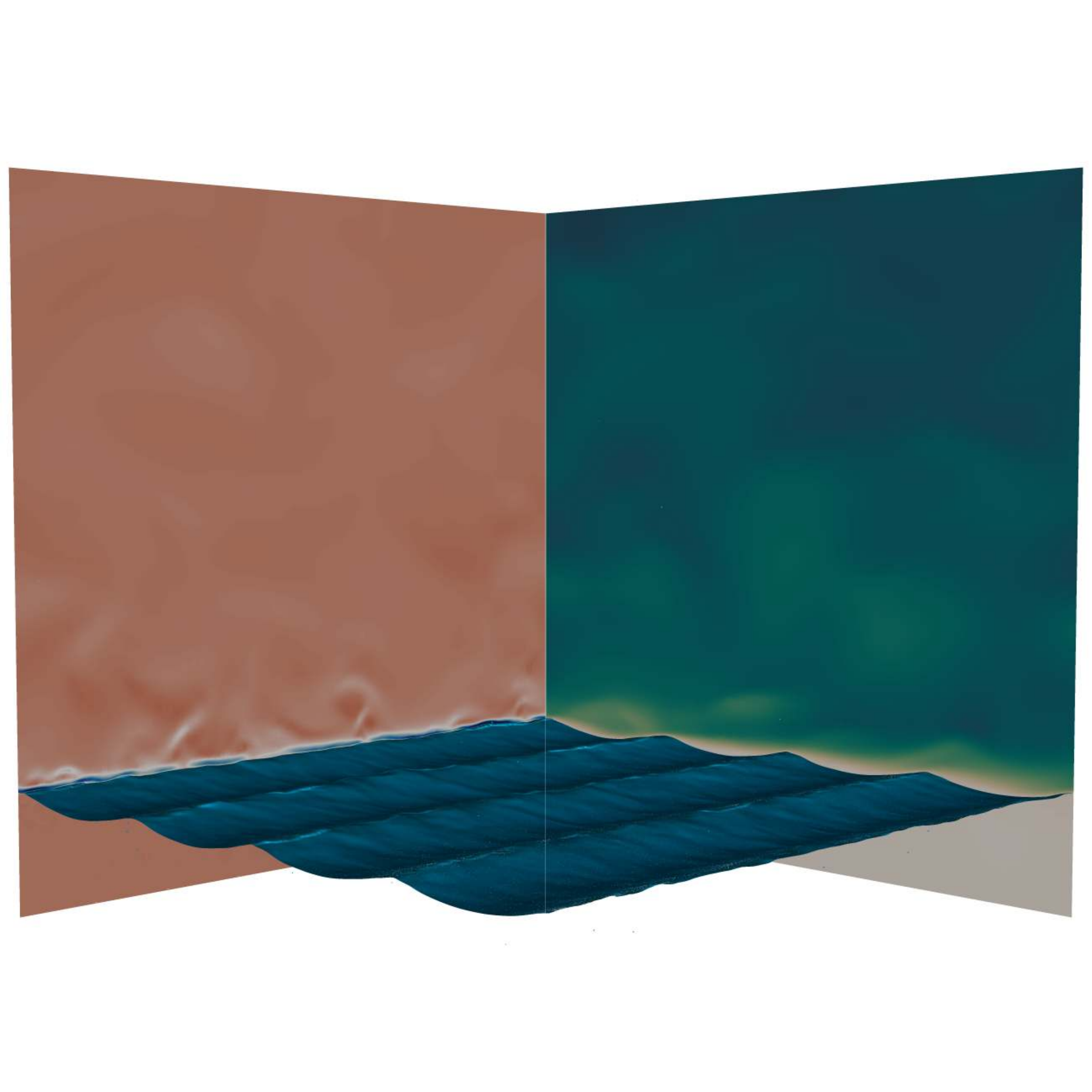}
  \hspace{0.25 cm}
  \includegraphics[trim={0cm 4.0cm 0cm 6.0cm}, clip, width=0.445\textwidth]{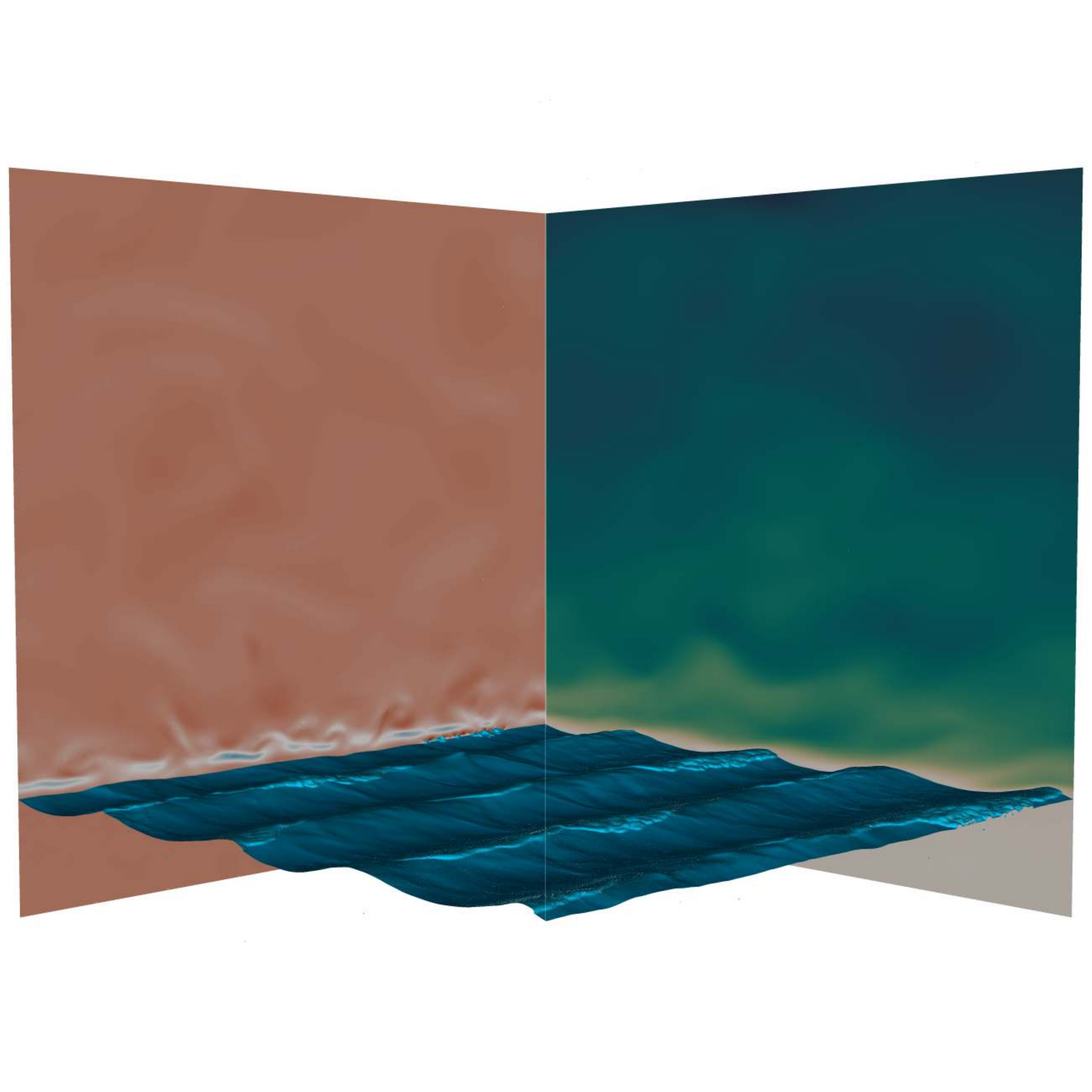}
  \put(-368,120){\small(c)}
  \put(-182,120){\small(d)} \\
  \includegraphics[width=0.45\textwidth]{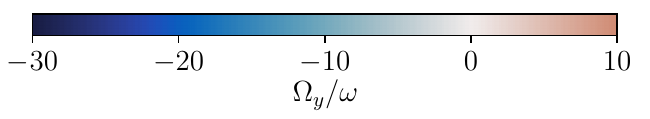}
  \hspace{0.25 cm}
  \includegraphics[width=0.45\textwidth]{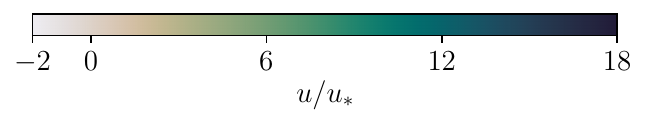}
  \caption{\revB{Wind-forced breaking waves for the case $u_\ast/c=0.9$ at $\omega t = 20-40-70-150$, (a, just before breaking)-(b, during breaking)-(c, during second growing stage)-(d, final stage with a more three-dimensional field and microbreaking), respectively. The turbulent airflow, i.e. the wind, and the waves move parallel along the positive streamwise direction, $x$}. In all the figures, the plane on the left contains the contour of the \revC{local spanwise vorticity $\Omega_y=\left(\partial w/\partial x-\partial u/\partial z\right)$} normalized by the wave angular velocity $\omega=2\pi/T_0$, the plane on the right contains the contour of the \revC{local} streamwise velocity $u$ normalized by the friction velocity $u_\ast$.}
  \label{fig:evol_the_system}
\end{figure*}

Figure~\ref{fig:evol_the_system} shows the wind-wave system for the case $u_\ast/c=0.9$ at four characteristic times. At $\omega t = 20$ (fig.~\ref{fig:evol_the_system}(a), the wave field has grown due to the wind input and is close to breaking conditions (incipient breaking). While the initial wave field is nearly monochromatic, the waves become sharp-crested as they grow under wind forcing and approach breaking. Once the breaking stage is concluded, around $\omega t=41$ (fig.~\ref{fig:evol_the_system}(b), the wave field grows again, starting with a smaller steepness (fig.~\ref{fig:evol_the_system}(c) taken at $\omega t = 70$ and finally breaks again at $\omega t = 150$ (fig.~\ref{fig:evol_the_system}(d). \revC{In this last condition, the wave field becomes clearly three-dimensional, with energy-redistribution to higher wavenumber following the breaking and growing cycles, and intermittent microbreaking events visible. Despite this, the ratio between the effective main wavelength and the initial wavelength remains nearly unchanged, with wave elevation analysis showing variations of less than 6\% throughout the simulations (not shown). Moreover, most of the wave energy remains concentrated around the peak wavelength set by the initial conditions, as constrained by the boundary conditions. The peak downshift is effectively limited by these constraints, with energy redistribution to higher wavenumbers primarily occurring following breaking events.} %However, energy re-distribution occurss during breaking with XX\% of the initial wave energy being transfered to high frequency modes. Need an analysis of the wave spectra for this.} 
\subsection{Time evolution of the potential wave energy}\label{sec:Erms_evol}
We consider the potential energy of the wave, $E_W(t)$, to characterize the evolution of a wave field and distinguish between the growing (i.e. $E_W$ increases) and the breaking stages (i.e. $E_W$ decreases),
\begin{equation}\label{eqn:E_W}
  E_W(t) = \rho_w|\mathbf{g}|\int_{\Omega_w} z dV - E_{p,0}\mathrm{,}
\end{equation}
where the integration volume in the water $\Omega_w$ is done up to the wave surface $\eta$ in the $z-$direction. $E_{p,0}$ is the potential energy when the wave field is undisturbed and can be evaluated as $E_{p,0}=\rho_w|\mathbf{g}|L_0^2\int_0^{h_W} zdz = \rho_w|\mathbf{g}|(h_WL_0)^2/2$. A contribution due to surface tension exists in $E_W(t)$ but is negligible here due to the large $Bo$. \revB{Note that we calculate $E_W(t)$ by integration in the water volume over the vertical coordinate $z$ \citep{lamb1993hydrodynamics} since it is also valid for non-linear and breaking waves contrary to the formula based on linear approximation $E_W(t)=\rho_w|\mathbf{g}|\int_\Gamma \sqrt{2(\eta-\overline{\eta})^2} dS$ \citep{lamb1993hydrodynamics} which is not well defined during the breaking stage.} \par
\begin{figure*}%[t!]
  \centering
  \includegraphics[width=0.53\textwidth, height=5 cm]{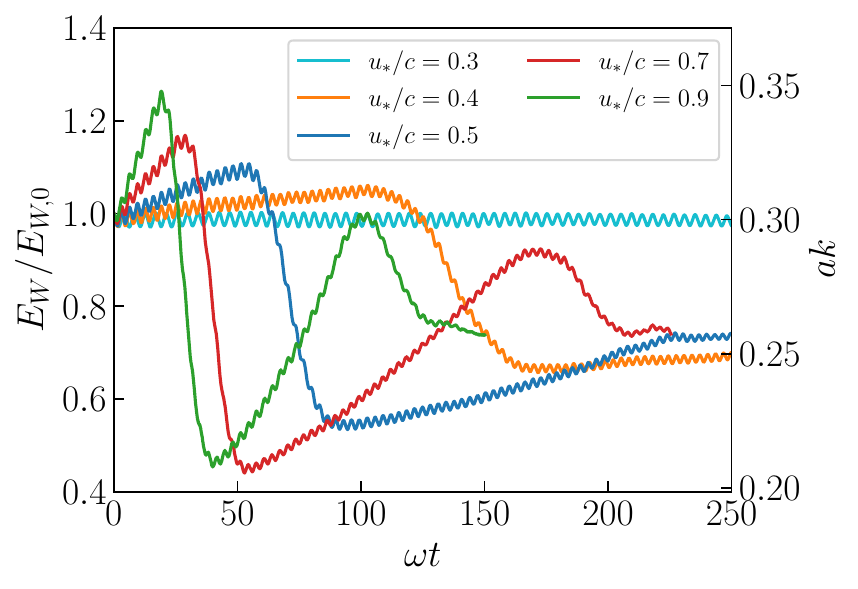}
  \includegraphics[width=0.45\textwidth, height=5 cm]{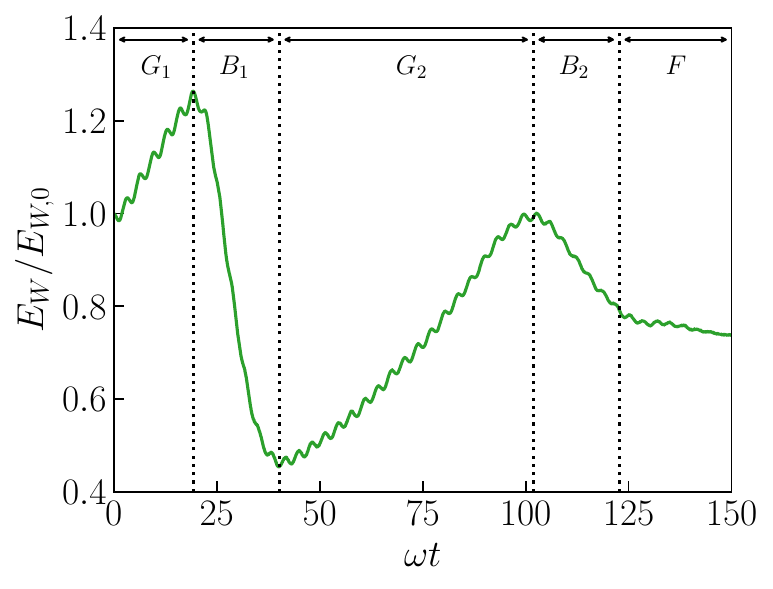}
  \put(-385,130){\small(a)}
  \put(-180,130){\small(b)}
  \caption{Wind-waves growth and breaking life cycle. (a) Evolution of the normalized potential wave energy $E_W/E_{W,0}$ with $E_{W,0}=E_W(t=0)$ for increasing $u_\ast/c$ from $0.3$ to $0.9$ as a function of the dimensionless time $\omega t$ where $\omega=2\pi/T_0$ is the angular frequency and $T_0$ is the wave period. \revC{The associated instantaneous steepness $ak$ values are shown on the second $y$-axis.} (b) Sketch of the characteristics of dynamical regimes observed in the simulations, illustrated for $u_\ast/c=0.9$. $G_{1,2}$ represent the first and the second growing stages, $B_{1,2}$ represent the first and the second breaking stages; $F$ is the final stage. $G_{2,a}$ and $G_{2,b}$ are the fractions of the second growing stages with equal time windows of the stages $G_1$ and $F$. These windows will be used to compute averages for the momentum flux and drag coefficient during growth and breaking.}
  \label{fig:E_rms}
\end{figure*}
%
%Figure~\ref{fig:E_rms}(a) shows the time evolution of the potential wave energy normalized by its initial value, $E_W/E_{W,0}$, (left axis), together with the steepness $\mathcal{S}$ (right axis, same lines), and illustrates the growing-breaking life cycle of the wave field under strong wind forcing. \revC{Note that as mentioned in}. \revC{The wave steepness Here $\mathcal{S}$ is generally evaluated from the derivatives of the surface elevation function as $\mathcal{S}=\max(\sqrt{(\partial\eta/\partial x)^2+(\partial\eta/\partial y)^2})$. It is worth mentioning that if linear wave theory is applicable, $\mathcal{S}$ reduces to $\mathcal{S}=a(t)k$ with $a(t)=\sqrt{(2/\Gamma)\int_\Gamma(\eta-\overline{\eta})^2dA}$. In the present work, both approaches have been tested and despite the assumptions the linear wave theory are not satisfied, they both provide very similar results.} \par
%
Figure~\ref{fig:E_rms}(a) depicts the time evolution of the potential wave energy normalized by its initial value, $E_W/E_{W,0}$ (left axis), highlighting the growth and breaking stages of the wave field under strong wind forcing. \revC{For reference, figure~\ref{fig:E_rms}(a) also presents the instantaneous wave steepness $ak$ (right axis), calculated from the wave amplitude as $a(t) = \sqrt{(2/\Gamma)\int_\Gamma (\eta - \overline{\eta})^2dS}$. Alternatively, the wave steepness can be determined using the surface elevation derivatives as $\mathcal{S} = \max(\sqrt{(\partial\eta/ \partial x)^2 + (\partial \eta/\partial y)^2})$. Note that under the assumption of linear wave theory, $\mathcal{S}$ reduces to $\mathcal{S} = a(t)k$. In this study, both methods were evaluated, and despite the assumptions of linear theory being unmet during breaking, they produced very similar results.} \par
The time evolution of wave energy $E_W/E_{W,0}$ and the wave steepness $ak$ is controlled by the competing effects of the wind forcing $u_\ast/c$ and the dissipation. \revC{In each curve, the potential wave energy displays small oscillations of period $T_0/2$, already observed in in~\citet{iafrati2009numerical,deike2015capillary} in similar configurations without wind.} 

During the growing stage, the dissipation is viscous and controlled by the water wave Reynolds number $Re_W$. \revB{For the smaller forcing $u_\ast/c=0.3$, the wind energy input and the viscous dissipation are in balance. Note that in ocean water wave, the wave Reynolds number is significantly higher and, therefore, the wave field would grow for $u_\ast/c=0.3$. Performing simulation at higher wave Reynolds number would also lead to wave energy growing in time for such value of $u_\ast/c=0.3$, see \citet{wu2021wind,wu2022revisiting,zhang2023wind}}. The wave growth rate increases with $u_\ast/c$ and for all stronger forcing, $E_W(t)$ grows until it reaches a critical point (breaking conditions), when part of the wave energy is dissipated. Once the breaking stage concludes, $E_W(t)$ grows again, and a second breaking event can occur at later times. When $u_\ast/c$ increases, the wave field reaches a critical amplitude for breaking $(ak)_c$ earlier. The observed critical breaking steepness $(ak)_c$ varies slightly from $(ak)_c\in [0.28-0.34]$ between the different wind forcing and first and second cycles, close to the values reported without wind for similar initial conditions~\citep{deike2015capillary,deike2016air} and experiments on focusing wave packets~\citep{drazen2008inertial}. The magnitude of the breaking event can be quantified by the maximum energy loss, $(\max(E_W)-\min(E_W))$, which increases with the amplitude of breaking. \par 
\revB{In all the cases, after one of two breaking events, the wave field reaches a final stage with limited variation in the wave energy. This condition can be attributed to the balance between the wind forcing and the wave field characterized by a broader wave spectrum and intermittent micro breaking events occurring at some of the wave crests (see figure~\ref{fig:evol_the_system}(d)). Such conditions reduce the effectiveness of wind input in promoting wave growth and lead to a quasi-steady condition with wind input approximately balanced by energy dissipation. Note that the loss of energy of these micro-breaking events is smaller since the steepness at breaking is smaller (see, for example, discussions in~\cite{drazen2008inertial,deike2015capillary}).} \par
For the purposes of the present analysis, we define different time windows that characterize the physics at play made of the growing and breaking stages, which are illustrated in figure~\ref{fig:E_rms}(b). We consider the first and second growing stages, $G_1$ and $G_2$ (when $E_W$ increases). Each growing stage is followed by a corresponding breaking stage, $B_1$ and $B_2$ (when $E_W$ decreases). Note that for the highest $u_\ast/c=0.7-0.9$, we run the simulations long enough to observe a breaking stage, followed by a quasi-stationary final state, $F$, where $E_W$ exhibits a limited variation. In the second growing stage, we define two further time windows $G_{2,a}$ and $G_{2,b}$. These time windows, together with $G_{1,2}$, $B_{1,2}$, and $F$, will be useful to define the mean velocity profiles, perform momentum flux and drag coefficient analysis during the growing and breaking stages and determine the respective role of each regime in the averaged quantities. Note that $G_1$-$G_{2,a}$ and $G_{2,b}$-$F$ are chosen to have the same duration in time, so that the averaged quantities defined therein are computed over comparable temporal windows.
\subsection{Airflow above growing and breaking waves}
\begin{figure*}%[t!]
  \centering
  \includegraphics[width=\textwidth]{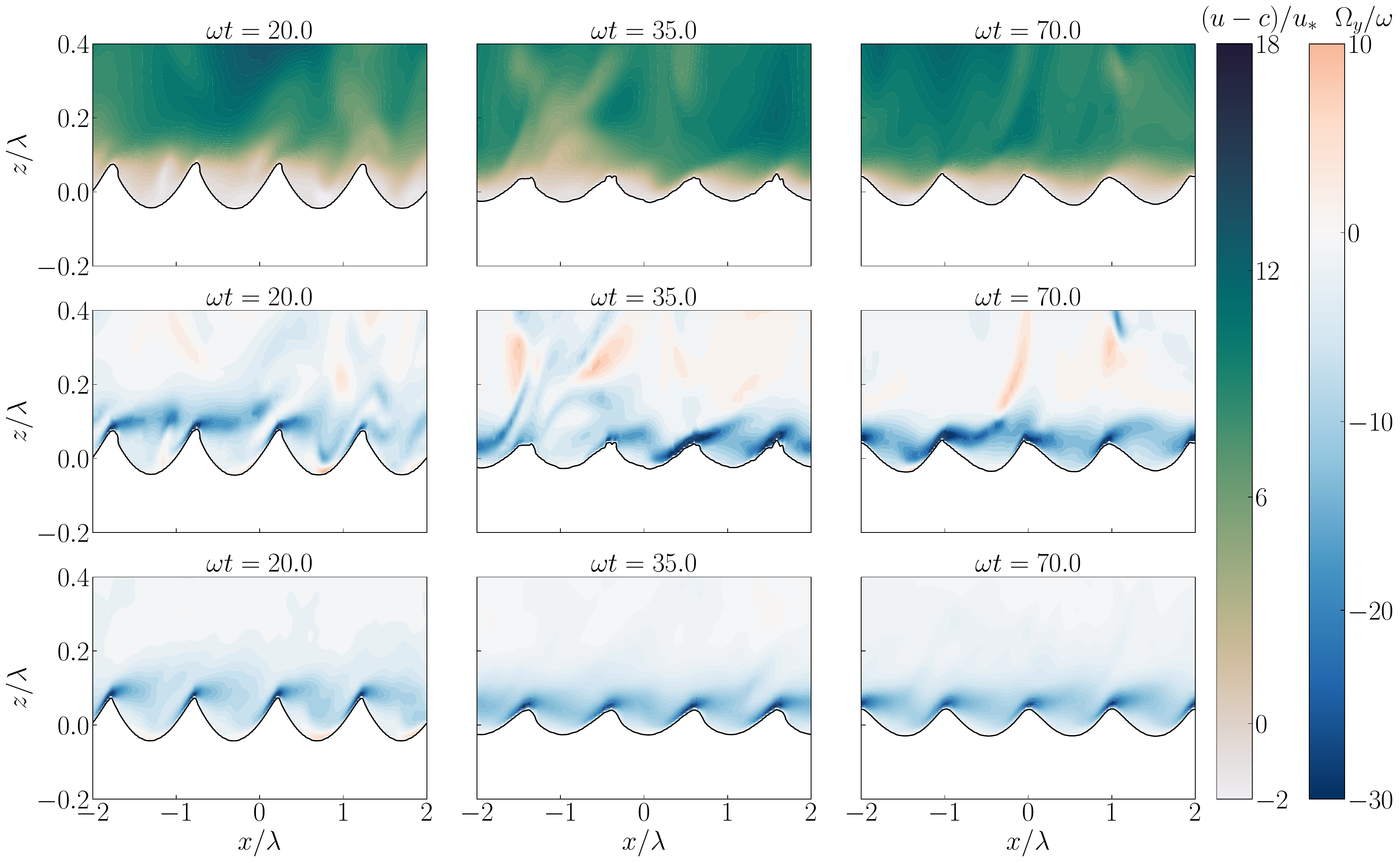}
  \put(-381,215){\small(a)}
  \put(-266,215){\small(b)} 
  \put(-159,215){\small(c)} 
  \put(-381,145){\small(d)}
  \put(-266,145){\small(e)} 
  \put(-159,145){\small(f)}
  \put(-381,072){\small(g)}
  \put(-266,072){\small(h)} 
  \put(-159,072){\small(i)}
  \caption{
  \revA{Illustration of air-flow separation in strongly forced steep waves, just before breaking (first column), during breaking (second column) and during the second growing stage (third column).
  Panels (a)-(b)-(c): contours of the streamwise instantaneous velocity in the airflow sampled at the middle plane $y/\lambda=0$. The streamwise velocity is plotted in a reference frame moving with the wave and is normalized by the friction velocity, i.e. $(u-c)/u_\ast$. 
  Panels (d)-(e)-(f): contours of the instantaneous spanwise vorticity in the airflow sampled at the middle plane $y/\lambda=0$ and normalized by the angular velocity, i.e. $\Omega_y/\omega$. 
  Panels (g)-(h)-(i): contours of the spanwise vorticity in the airflow, averaged along the spanwise direction and normalized by the angular velocity, i.e. $\Omega_y/\omega$. 
  All the panels are plotted in the region $-0.2\lambda\leq z \leq 0.4\lambda$ for $\omega t=[20,35,70]$ for $u_\ast/c=0.9$.}
  }
  \label{fig:flow_sep}
\end{figure*}
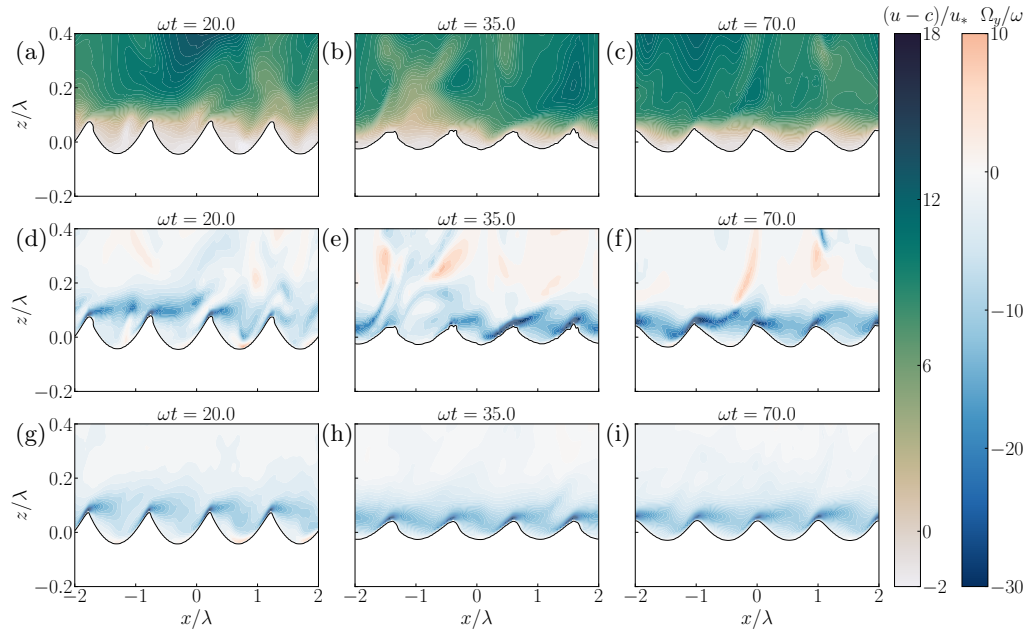
\revA{
The flow structure near the interface is heavily modified by the presence of the waves. Figure~\ref{fig:flow_sep}(a)-(b)-(c) shows the instantaneous velocity contours of the streamwise air velocity in a reference frame moving at the wave speed, i.e. $u-c$, in the region close to the wave surface ($-0.2\leq z\leq 0.4\lambda$) and sampled at the spanwise mid-plane ($y/\lambda=0$), at several characteristic times (during growth and breaking) for $u_\ast/c=0.9$. \par
%
%During the initial growing stage, $u-c$ is very close to zero or slightly below zero in some wave troughs, as shown in panel (a). As the wave field grows and approaches breaking, i.e. $\omega t=20$ in panel (b), large negative values appear and localize not only near the wave troughs but also close to the wave crests. In these regions, the flow separates and recirculates, with negative velocity up to $1.5u_\ast$. Note that flow separation starts to occur before the breaking stage, as the wave becomes steep and short-crested. During the post-breaking stage for $\omega t>35$, when the wave amplitude is greatly reduced, the regions with negative $u_a-c$ reduce compared to the growing stage, see panels (c,d). \par
%
As the wave field grows and approaches breaking, i.e. $\omega t=20$ in panel (a), large negative values appear and localize not only near the wave troughs but also close to the wave crests. In these regions, the flow reverses and recirculates, with negative velocity up to $1.5u_\ast$. Note that flow reversal starts to occur before the breaking stage, as the wave becomes steep and short-crested. During the post-breaking stage for $\omega t>35$, when the wave amplitude is greatly reduced, the regions with negative $u-c$ reduce compared to the growing stage, see panels (b,c). \par
As noted in~\citep{veron2007measurements}, although flow reversal suggests airflow separation from the wave field, a more rigorous approach involves examining the spanwise vorticity $\Omega_y$, which is a Galilean-invariant quantity. The spanwise vorticity normalized by the angular velocity, i.e. $\Omega_y/\omega$, is reported in panels (d)-(e)-(f) of figure~\ref{fig:flow_sep}. For the different panels, a thin layer of vorticity is observed to form on the windward side of the wave, i.e. before the wave crest, where the turbulent airflow shears this layer, causing it to detach from the surface and mix with the flow. Lacking support from the wave field, the detached vorticity "packet" destabilizes, breaking down before gradually reforming at the next wave crest. Although the panels (d)-(e)-(f) refer to a specific region of the domain, i.e., $y/\lambda=0$, flow separation occurs all over the wave crests, as shown in panels (g)-(h)-(i), where the spanwise-averaged vorticity is displayed. This observation suggests that, in the present configuration with steep waves under breaking conditions, flow separation occurs over all the wave crests, as also observed in~\citep{buckley2020surface}. \par
}
Flow separation is an important feature of wind-forced breaking waves and has been argued to enhance the momentum flux in breaking waves~\citep{banner1976separation,reul648air,buckley2020surface}, and is analyzed in the next section. 
%
% Momentum flux
%
\section{Momentum fluxes over strongly forced breaking waves}\label{sec:mom_flux}
We now analyze the momentum fluxes between the turbulent airflow and the growing and breaking waves. In \S~\ref{sec:mom_bud}, we examine the partition of the total momentum flux into pressure and viscous forces and their temporal variation during a breaking cycle. In \S~\ref{sec:vel_prof}, we investigate how the velocity profile and the Reynolds stress of the airflow are modulated by waves during the growing and breaking stages. 
\subsection{Momentum budget}\label{sec:mom_bud}
The momentum flux, $\boldsymbol\tau_{t}=(\tau_{t,x},\tau_{t,y},\tau_{t,z})$, is defined as the total force exerted from the wind to the waves per unit of air mass density and can be decomposed in pressure and a viscous contribution~\citep{belcher1998turbulent,sullivan2000simulation}. For water waves moving parallel to the wind in the streamwise $x$ direction, the first force component, i.e. $\tau_{t,x}$, is typically dominant and reads
\begin{equation}
  \tau_{t,x} = \tau_{p,x} + \tau_{\nu,x} = - \int_\Gamma p\mathbf{n}\cdot\mathbf{e}_xdS + 2\mu_a\int_\Gamma(\mathbf{D}\mathbf{n})\cdot\mathbf{e}_xdS\mathrm{,}
  \label{eqn:Ftx}
\end{equation}
where both the pressure $p$ and \revB{the strain rate tensor $\mathbf{D}$} are evaluated in the air phase. The two contributions to the right-hand side of equation~\eqref{eqn:Ftx} are termed pressure (or form) drag and viscous drag. \par
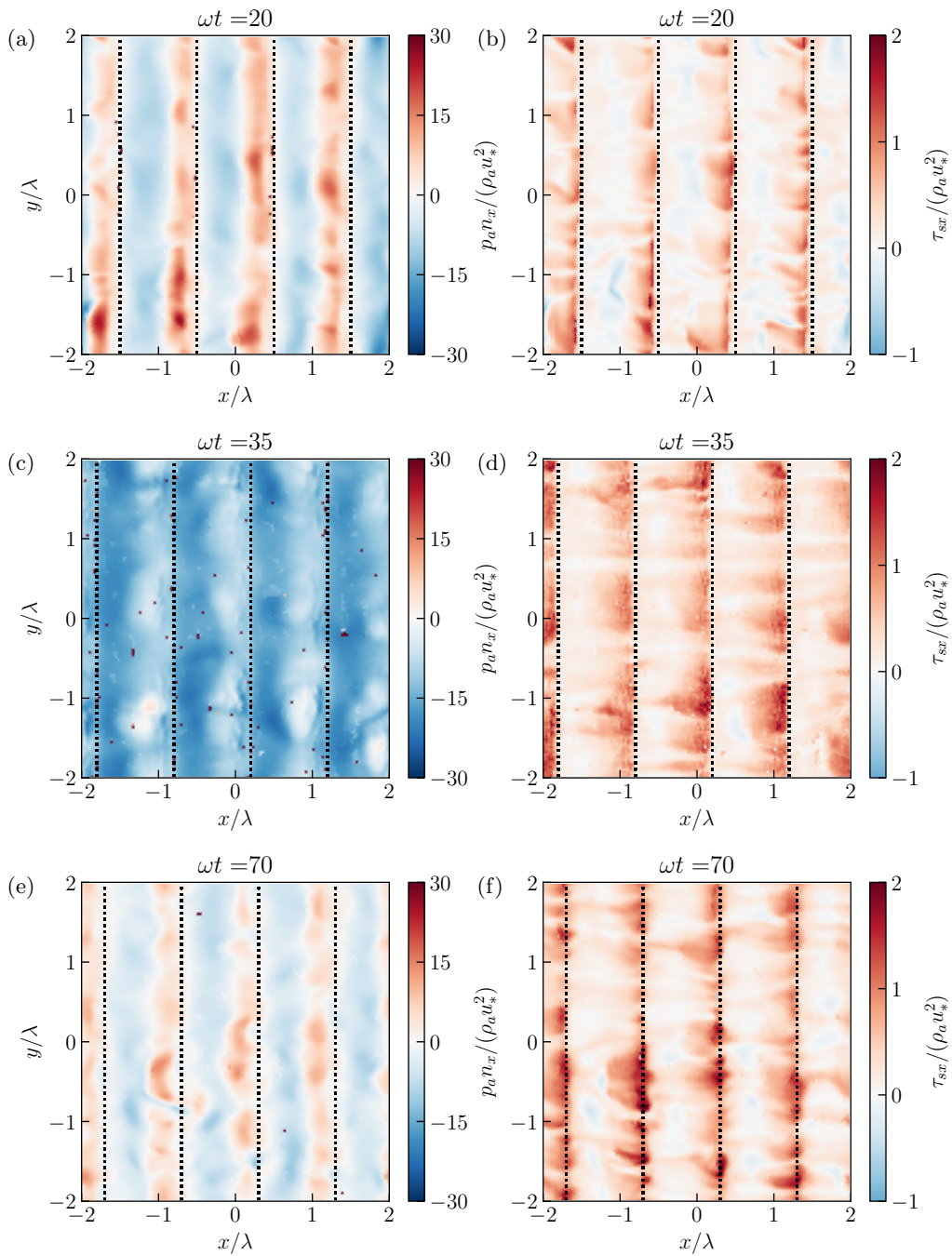
\begin{figure*}%[t!]
  \centering
  \includegraphics[width=\textwidth]{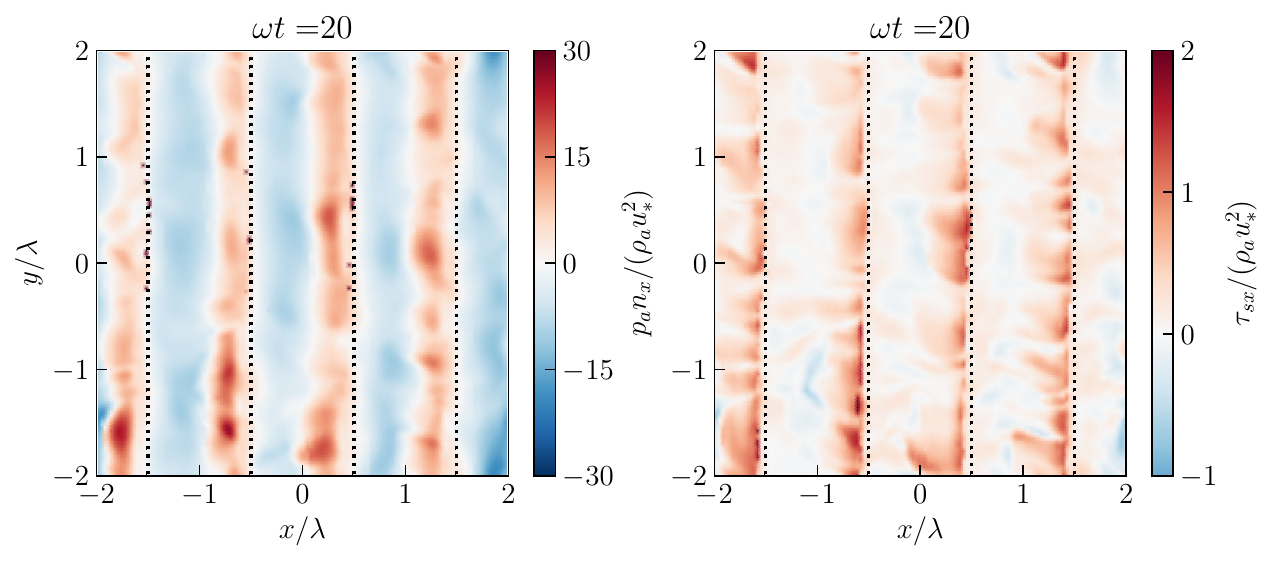}
  \put(-385,150){\small(a)}
  \put(-195,150){\small(b)} \\
  \includegraphics[width=\textwidth]{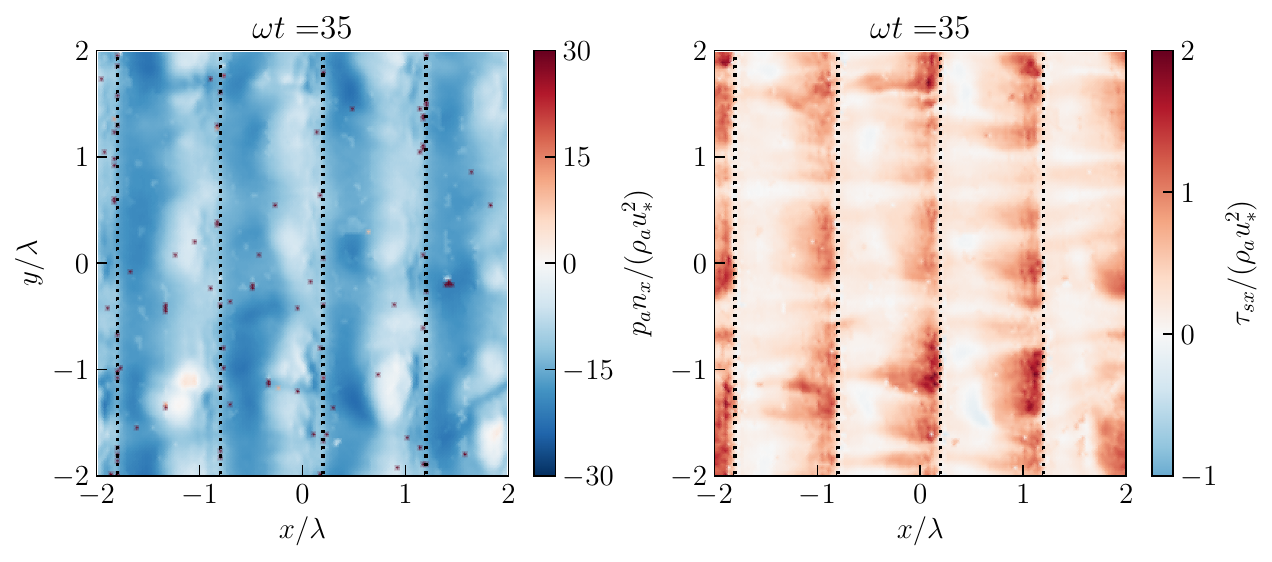}
  \put(-385,150){\small(c)}
  \put(-195,150){\small(d)} \\
  \includegraphics[width=\textwidth]{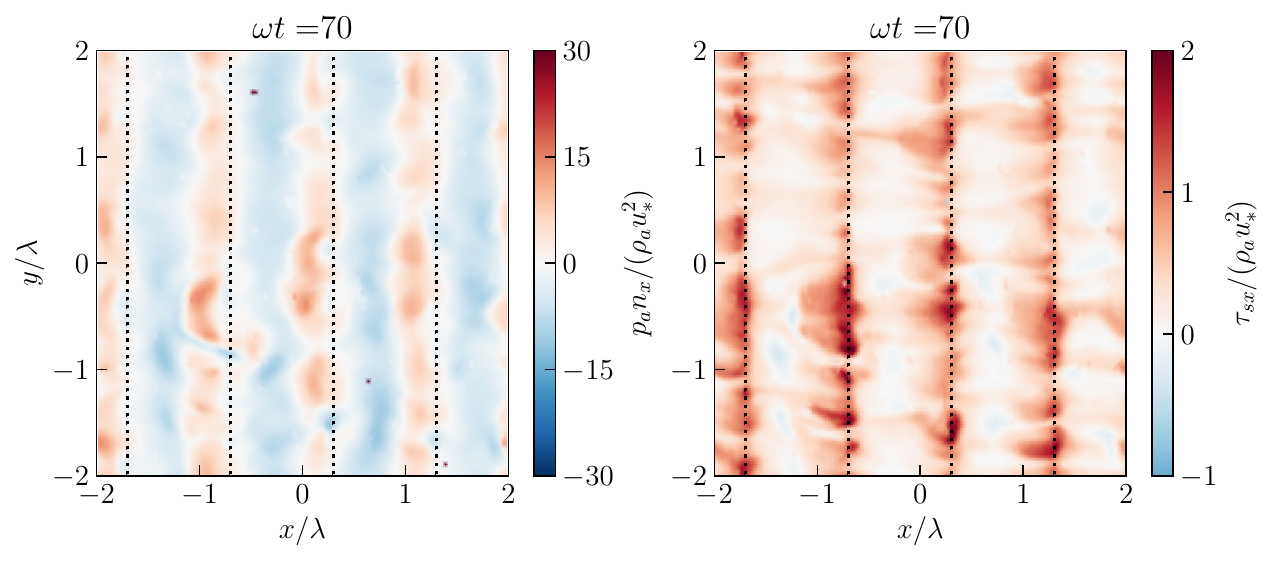}
  \put(-385,150){\small(e)}
  \put(-195,150){\small(f)} %\\
  \caption{Surface distribution of the pressure stress $p_an_x$ (panels $\mathrm{(a)}-\mathrm{(c)}-\mathrm{(e)}$) and viscous $\tau_{sx}=2\mu_a\left(\mathbf{D}\mathbf{n}\right)\cdot\mathbf{e}_x$ stress (panels $\mathrm{(b)}-\mathrm{(d)}-\mathrm{(f)}$) for $u_\ast/c=0.9$. Note that both quantities are normalized by the total imposed stress $\rho_au_\ast^2$ at the interface. The dot-dashed lines in all the panels represent the position of the wave crest.}
  \label{fig:ptau_dist_0p9}
\end{figure*}

We start by qualitatively inspecting both contributions during the growing and breaking stages. Figure~\ref{fig:ptau_dist_0p9} reports the instantaneous distribution of $p_an_x$ and $2\mu_a\left(\mathbf{D}\mathbf{n}\right)\cdot\mathbf{e}_x$ for the case $u_\ast/c=0.9$ at three physical times $\omega t=[20, 35, 70]$, corresponding to the growing pre-breaking stage, breaking stage, and to the second growth post-breaking. \revC{Hereinafter, we define $\tau_{sx}=2\mu_a\left(\mathbf{D}\mathbf{n}\right)\cdot\mathbf{e}_x$ for brevity and, therefore,  $\tau_{\nu,x}=\int_\Gamma \tau_{sx}dS$ in equation~\eqref{eqn:Ftx}}. The distributions $p_an_x$ and $\tau_{sx}$ are both projected on a wave-following surface, very close to the surface (vertically shifted by $0.1/k$ from the wave surface, see~\cite{wu2022revisiting} and appendix~\ref{app:fluxes} for details). \par
Figure~\ref{fig:ptau_dist_0p9} shows that both $p_an_x$ and $\tau_{sx}$ display clear wave-induced coherent patterns, together with the development of three-dimensional structures induced by turbulence. When the wave field grows, i.e. panels (a-b) and (e-f), \revC{the pressure assumes negative values in the wave troughs and positive values near the wave crest. Similarly, the viscous stress also assumes positive value near the wave crest, whereas it approaches zero near the wave troughs and intermittently becomes negative at $\omega t=20-70$, consistently with the measurements in~\citep{veron2007measurements}}. Furthermore, the peaks in pressure appear on the windward face (on the left of the dotted black lines at the wave crest), whereas the peak in the viscous stress is localized near the wave crest, mainly due to turbulence-induced straining of the shear layer. Note that the peaks of $pn_x$ are one order of magnitude larger than the peak in $\tau_{sx}$. This result is consistent with the large initial slope of the waves for which the pressure force is the leading term in equation~\eqref{eqn:Ftx}, as also shown in experimental works~\citep{buckley2020surface}. When the wave breaks, i.e. panels (c-d), the distribution of $pn_x$ changes dramatically with a loss of the wave-coherent pattern composed of high-pressure and low-pressure regions in the windward and leeward sides, respectively. Conversely, the viscous stress distribution is much less affected by breaking. \par
We can quantify how the abrupt change in the form drag distribution reflects in the total momentum budget.
We integrate the streamwise component of the momentum equation~\eqref{eqn:mom_con} over the air volume $\Omega_a$:
\begin{equation}
  \underbrace{\rho_a\dfrac{\partial}{\partial t}\int_{\Omega_a} udV}_{\rho_a\partial U/\partial t} + \underbrace{\rho_a\int_{\Gamma}u(\mathbf{u}_r\cdot\mathbf{n})dS}_{\rho_a\phi_{c,x}} = \tau_{p,x} + \tau_{\nu,x} + \underbrace{\rho_au_\ast^2A_\Gamma}_{\Pi_f}\mathrm{,}
  \label{eqn:u_a}
\end{equation} 
where $dV$ is the elementary volume in the airflow region, $\Pi_f$ is the volume-integrated imposed pressure gradient (defined in~\eqref{eqn:pif}), i.e. $\Pi_f=\int_{\Omega_a}\partial p_0/\partial x (1-\mathcal{F})dV=\rho_au_\ast^2A_\Gamma$ with $A_\Gamma=\Omega_a/(L_0-h_W)$. In equation~\eqref{eqn:u_a}, the left-hand side is the total rate of change in the air velocity, accounting for the temporal variation $\rho_a\partial U/\partial t$ and the convective contribution $\rho_a\phi_{c,x}$. The former term, $\rho_a\partial U/\partial t$, accounts for the response of the instantaneous flow field to the variation in the momentum total flux $\tau_{t,x}=\tau_{p,x}+\tau_{\nu,x}$, and the latter, $\rho_a\phi_{c,x}$, accounts for the momentum flux originated from a non-zero relative velocity $\mathbf{u}_r$ between the airflow and the wave field. Since the magnitude of $\mathbf{u}_r$ is enhanced in the presence of airflow separation events, the term $\rho_a\phi_{c,x}$ can be employed to quantify such events over growing and breaking waves. The right-hand side includes the uniform forcing term $\Pi_f$ and the two forces per unit of mass $\tau_{p,x}$ and $\tau_{\nu,x}$, as in equation~\eqref{eqn:Ftx}. Note that in a statistically steady condition with negligible relative velocity between the airflow and the wave field, the two terms on the left-hand side of~\eqref{eqn:u_a} vanish, i.e. $\langle\rho_a\partial U/\partial t\rangle_t=\langle\rho_a\phi_{c,x}\rangle_t=0$ (with $\langle\rangle_t$ a time-averaged operator). Accordingly, equation~\eqref{eqn:u_a} reduces to the simplified form where the total imposed stress at the interface $\rho_au_\ast^2A_\Gamma$ balances the sum of pressure and viscous drag~\citep{janssen2004interaction,buckley2020surface,funke2021pressure}. More details on the evaluation of the momentum fluxes are provided in the Appendix~\ref{app:fluxes}. \par
\begin{figure*}%[t!]
  \centering
  \includegraphics[width=\textwidth]{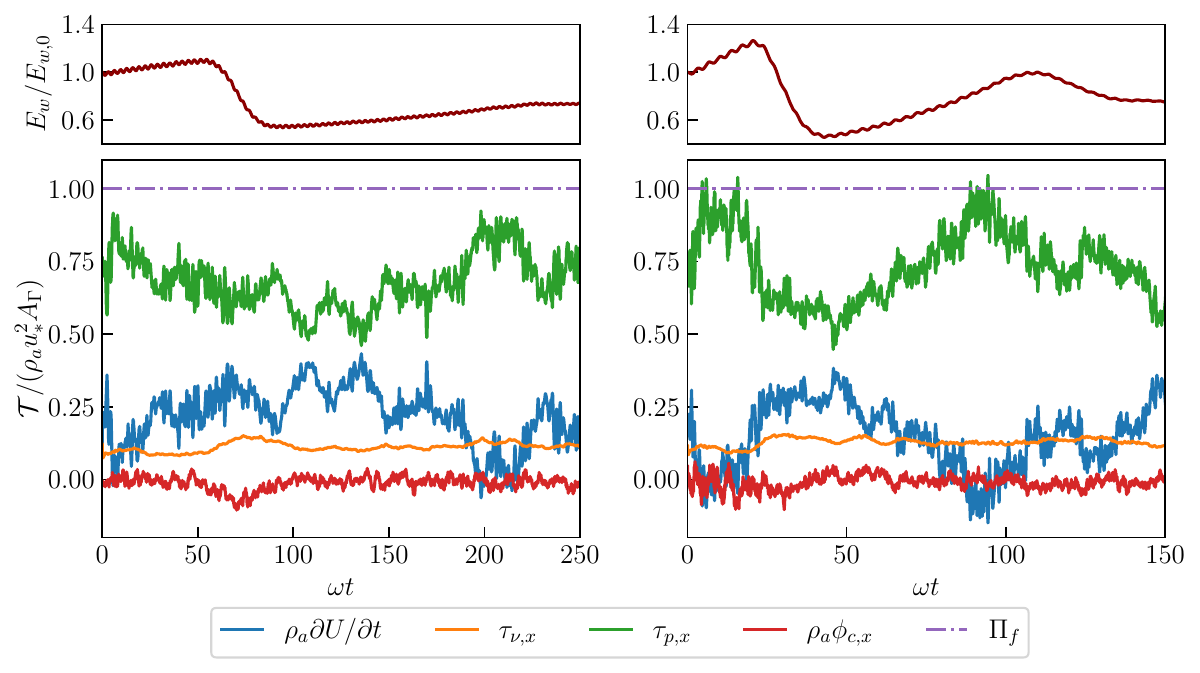}
  \put(-385,150){\small(a)}
  \put(-195,150){\small(b)}
  \caption{Contributions of the momentum budget in the streamwise direction, as in eq.~\eqref{eqn:u_a}, for: (a) $u_\ast/c=0.5$ and (b) $u_\ast/c=0.9$. On the $y$-label $\mathcal{T}$ represents the variation in the instantaneous flow $\rho_a\partial U/\partial t$, the viscous stress $\tau_{\nu,x}$, the pressure stress $\tau_{p,x}$, the convective term $\rho_a\phi_{c,x}$ or the driving force $\Pi_f$ (defined in equation~\eqref{eqn:pif}). Each budget component is normalized by the total stress $\rho_au_\ast^2A_\Gamma$. For both cases, the normalized variation in the wave energy $E_W/E_{W,0}$ is reported in top panel.}
  \label{fig:mom_flux_0p5_0p9}
\end{figure*}
We analyze the temporal variation of the different terms in the momentum flux eq.~\eqref{eqn:u_a} for two representative cases ($u_\ast/c=0.5-0.9$), in figure~\ref{fig:mom_flux_0p5_0p9}. The top panels display the normalized wave energy variation $E_W(t)$ to remind the reader of the overall time evolution. In both cases, the wave field experiences a first growth up to the breaking event ($\omega t\in [0-65]$, for $u_\ast/c=0.5$ and $[0-20]$ for $u_\ast/c=0.9$,  the shorter growing cycle for $u_\ast/c=0.9$ being due to the higher wind forcing). \par
When the wave breaks, the momentum flux due to pressure $\tau_{p,x}/(\rho_au_\ast^2A_\Gamma)$ drops with a corresponding acceleration of the flow, i.e. $\rho_a\partial U/\partial t/(\rho_au_\ast^2A_\Gamma)$ increases, the drop being larger for the strongest wind forcing. Once the breaking stage ends, around $\omega t=75$ for $u_\ast/c=0.5$ and around $\omega t=35$ for $u_\ast/c=0.9$, the wave field experiences a second growing cycle. During the interval, the pressure force remains the dominant contributor to the momentum budget and keeps increasing with a progressive deceleration of the airflow. For the case $u_\ast/c=0.9$, the wind forcing drives the wave field to a second breaking event, which occurs around $\omega t=120$. Similarly to the first breaking event, the pressure force decreases while the airflow accelerates, with the magnitude of the momentum loss being smaller (since the breaking is weaker). \par
In all cases, the momentum flux originated from the non-zero velocity between the airflow and the waves, $\rho_a\phi_{c,x}$, represents a negligible part of the momentum budget. This is consistent with \citet{yang2018direct}, which found a negligible role of the convective term in the budget. Next, the viscous contribution $\tau_{\nu,x}$ represents a small but not negligible contribution in the forces $\tau_{\nu,x}$. This is consistent with the high initial wave steepness, i.e. $a_0k=0.3$. Notably, upon wave breaking, the viscous contribution experiences a slight increase to partially compensate for the loss of the pressure force. \par
The sensitivity of the pressure force to the breaking event can be qualitatively understood by considering the variations of the instantaneous slope. When the wave breaks, the instantaneous wave slope, $\partial\eta/\partial x$, suddenly reduces and directly influences the dominant pressure stress $\tau_{p,x}$. (which is directly visible in a simplified version of eq.~\eqref{eqn:Ftx}, see \citet{funke2021pressure}). 
\subsection{Velocity profiles}\label{sec:vel_prof}
In \S~\ref{sec:mom_bud}, we showed that when the wave breaks, the momentum flux associated with pressure is reduced in favour of a flow acceleration in the air. We now characterize the associated streamwise vertical velocity profiles $\langle u_a\rangle$ during the growing-breaking cycle. \par
Following ~\citet{sullivan2000simulation,wu2022revisiting}, the mean velocity profile is computed in wave-following coordinates by using an implicit transformation, which maps the Cartesian $(x,y,z)$ to wave-following coordinates $(\xi,\eta,\zeta)$. This step allows us to perform the space average of the velocity field along the periodic directions by including the region below the wave crests. \revC{More details about the procedure to transform a generic field defined in a Cartesian coordinate system into a wave-following one are given in the appendix~\ref{app:map_cart}.} 
%
% we use~\citep{wu2022revisiting} 
%
%\begin{equation}\label{eqn:wave_fol}
%  \begin{bmatrix}
%    x \\
%    y \\
%    z
%  \end{bmatrix}
%    = \begin{bmatrix}
%    x(\xi,\zeta) \\
%    y(\eta) \\
%    z(\xi,\zeta)
%  \end{bmatrix}
%    = \begin{bmatrix}
%    \xi \\
%    \eta \\
%    \zeta + \overline{\eta}(\xi)\exp(-k|\zeta|)
%  \end{bmatrix}\mathrm{.}
%\end{equation}
%
%where $\overline{\eta}(\xi)$ is the spanwise-averaged surface elevation function. Note that for a sufficiently large value of $k|\zeta|$, $\exp(-k|\zeta|)\approx 0$ and the vertical Cartesian coordinate coincides with the vertical wave-following coordinate.} \par
%
\begin{figure*}%[t!]
  \centering
  \includegraphics[width=0.49\textwidth]{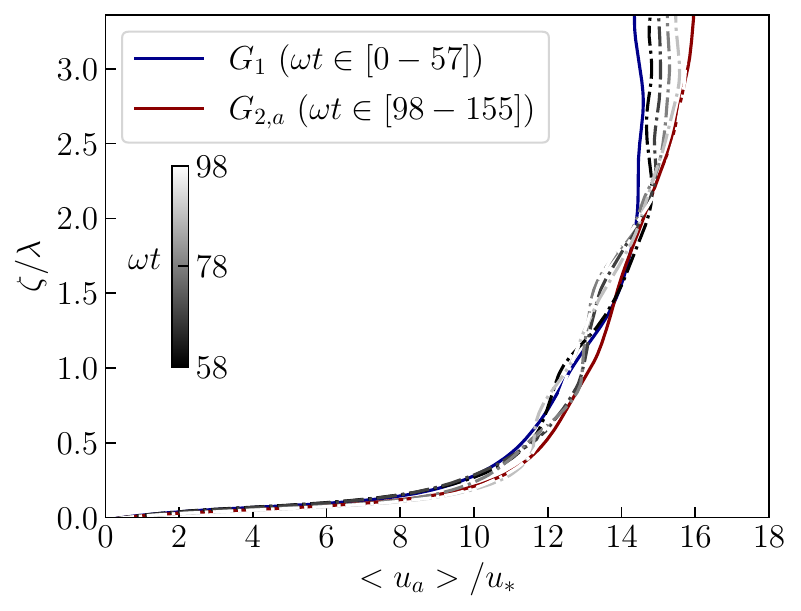}
  \includegraphics[width=0.49\textwidth]{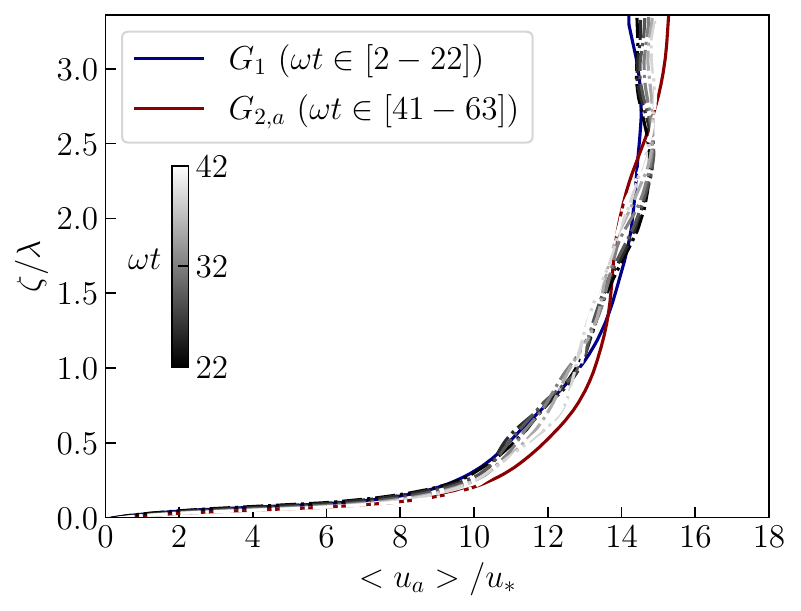}
  \put(-385,130){\small(a)}
  \put(-195,130){\small(b)}
  \caption{Streamwise velocity profile normalized by the nominal friction, $\langle u_a\rangle/u_\ast $, as a function of vertical wave-following coordinate $\zeta/\lambda$ in the airflow for: $(\mathrm{a})$ $u_\ast/c=0.5$ and $(\mathrm{b})$ $u_\ast/c=0.9$. For large enough values, $\zeta=z$. The instantaneous velocity profiles are averaged in time over the cycle $G_1$ (darkblue lines) and a fraction of the second growing cycle, $G_{2,a}$, as defined in section~\ref{sec:Erms_evol}, see figure~\ref{fig:E_rms}(b). The dot-dashed curves represent the instantaneous values during the breaking stage, i.e. $\omega t\in [58-98]$ for $u_\ast/c=0.5$ and $\omega t\in [22-42]$ for $u_\ast/c=0.9$.}
  \label{fig:profiles_0p5_0p9}
\end{figure*}
Figure~\ref{fig:profiles_0p5_0p9} shows the streamwise velocity profile spatially averaged along the horizontal directions as a function of the vertical wave-following coordinate $\zeta$ for $u_\ast/c=0.5-0.9$. In both cases, the velocity profiles are time-averaged over three time-windows. The two windows correspond to the two growing cycles, $G_1$ and $G_{2,a}$, introduced in figure~\ref{fig:E_rms}(b). \revA{Furthermore, note that the velocity profiles in the airflow are not subtracted by the streamwise component of the surface water velocity, as its maximum value in the water throughout the simulation is of the order of $0.33u_\ast$, which is negligible compared to the maximum streamwise velocity in the airflow (see figure~\ref{fig:profiles_0p5_0p9}).} \par %The second window corresponds to one physical time instant in the middle of the breaking stage. \par
By comparing the mean velocity profile $\langle u_a\rangle$ during the stages $G_1$ and $G_{2,a}$, we clearly see that the breaking event leads to a flow acceleration mainly confined in the region $\zeta/\lambda < 1$ and a shift of the velocity profiles to larger values, as also clearly shown from the instantaneous profiles during the breaking stage. The flow acceleration becomes more pronounced as we compare the case $u_\ast/c=0.5$ with the case $u_\ast/c=0.9$. The profile is non-zero at the water surface due to the presence of a developing underwater current. \par
Note that even during the growing stage, the simulation remains transient in nature. However, the turbulent flow adjusts to the moving wave field on a timescale $t_t=\nu_a/u_\ast^2$ that is $40$ to $200$ times smaller than the wave period $T_0 = \lambda/c$ for the different analyzed cases, i.e. $T_0/t_t = (u_\ast/c)Re_{\ast,\lambda} > 40-200$. Therefore, except during the breaking stage, the configuration can be approximated as quasi-stationary, allowing us to analyze the modulation of the velocity profile by the growing waves and the breaking event. \par
This analysis demonstrates the specific contribution of breaking events to the average velocity profiles, which, in laboratory or open ocean conditions, are typically averaged over long time periods without distinguishing between breaking and growing stages.
\subsection{Reynolds shear stress}\label{sec:turb_int}
\begin{figure*}%[t!]
  \centering
  \includegraphics[width=0.49\textwidth]{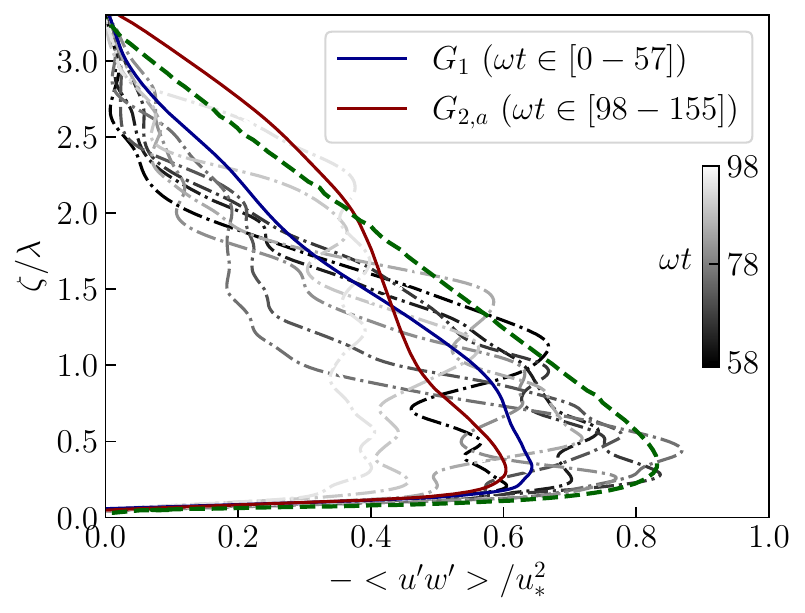}
  \includegraphics[width=0.49\textwidth]{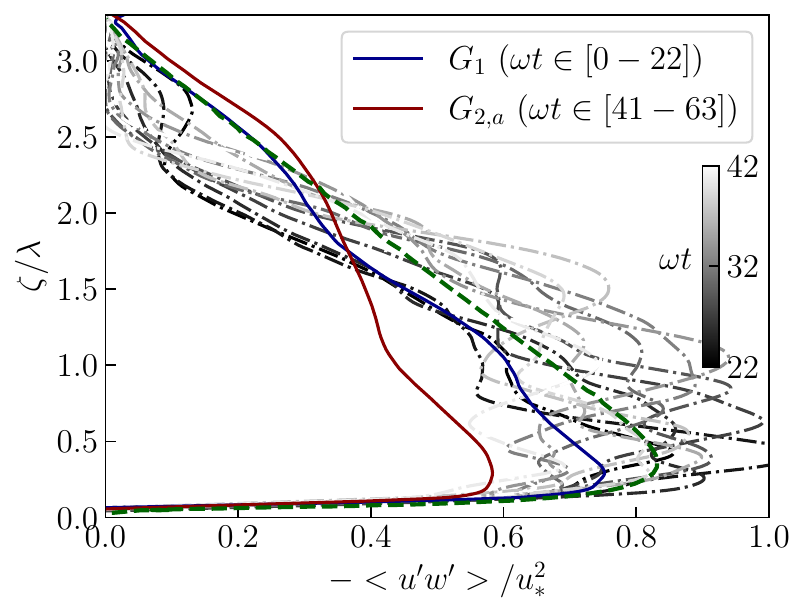}
  \put(-385,130){\small(a)}
  \put(-190,130){\small(b)}
  \caption{Reynolds shear stress normalized by the square of the nominal friction, $-\langle u'w'\rangle/u_\ast^2$, as a function of the vertical wave-following coordinate $\zeta$ for: (a) $u_\ast/c=0.5$ and (b) $u_\ast/c=0.9$. The Reynolds shear stress is averaged over the same time windows as in figure~\ref{fig:profiles_0p5_0p9}. The dot-dashed curves represent the instantaneous values during the breaking stage, i.e. $\omega t\in [58-98]$ for $u_\ast/c=0.5$ and $\omega t\in [22-42]$ for $u_\ast/c=0.9$. \revA{The dashed green curve represents the Reynolds stress on a flat stationary surface at $Re_\ast=720$}.}
  \label{fig:upwp_0p5_0p9}
\end{figure*}

The growing and breaking cycle also affects the Reynolds shear stress $\langle u'w'\rangle$ (or turbulent intensity) in the near wave-field region. Figure~\ref{fig:upwp_0p5_0p9} displays the Reynolds stress spatially averaged over the two periodic directions and normalized by $u_\ast^2$, i.e. $-\langle u'w'\rangle/u_\ast^2$, for $u_\ast/c=0.5-0.9$ (using the wave following coordinate as detailed in the Appendix~\ref{app:map_cart}). Each instantaneous profile is also time-averaged during the first and a fraction of the second growing cycle, i.e. $G_1$ and $G_{2,a}$. Note that we also report the different instantaneous values during the breaking stage of each case. \par
For both cases, $-\langle u'w'\rangle/u_\ast^2$ shows an increasing trend in the near-wave region, reaching a peak at \revA{$\zeta/\lambda\approx 0.26$, (i.e. $z/\lambda\approx 0.38$)}. Beyond this point, \revA{it decreases and approaches the dashed green curve in figure~\ref{fig:upwp_0p5_0p9}, which represents the Reynolds stress for a flat and stationary wall at $Re_\ast=720$~\citep{Pope_2000}}. However, it does not reach this solution exactly due to the non-stationarity of the flow field induced by the moving waves, i.e. growing and breaking stages. \par
During the growing stage, the peak in the Reynolds stress is $0.62$ for $u_\ast/c=0.5$, whereas it is $0.76$ for $u_\ast/c=0.9$. This larger peak is due to the increased drag, primarily from the pressure component, at higher $u_\ast/c$. When the wave field breaks, as shown from the dot-dashed curves, both peaks decrease without a significant change in the peak location, with a more pronounced reduction at larger $u_\ast/c$ related to the stronger breaking event. \par
Thus, wind-induced wave breaking greatly modulates the Reynolds stress and confirms the discussion made when analysing the pressure field. During the growing stage, the turbulent drag, i.e. $-\langle u'w'\rangle/u_\ast^2$, is larger as $u_\ast/c$ increases, while when the wave breaks, the turbulent drag is reduced, and such loss increases with $u_\ast/c$. We note that both cases are run at the same $Re_{\ast,\lambda}$, so the phenomenology is indeed controlled by the wind forcing and breaking dynamics and is much less affected by the Reynolds number.
\subsection{Extraction of the surface roughness from the velocity profiles}
\begin{figure*}%[t!]
  \centering
  \includegraphics[width=0.49\textwidth]{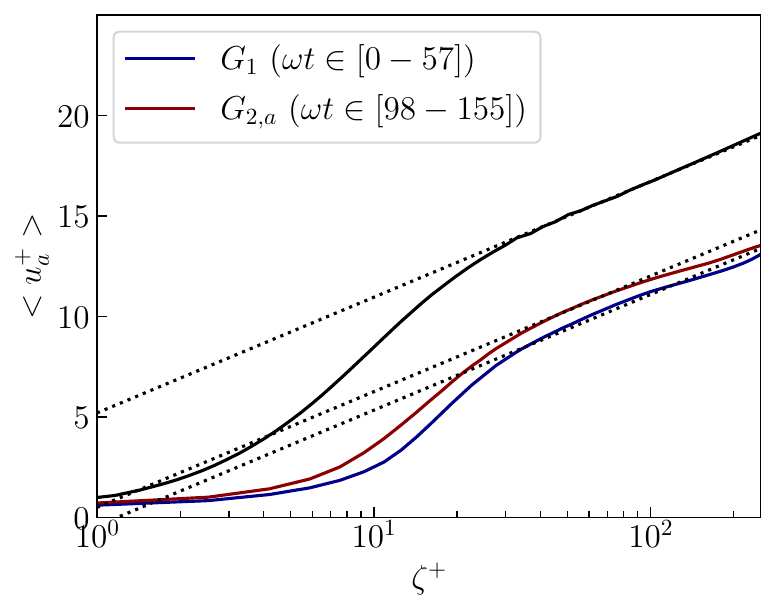}
  \includegraphics[width=0.49\textwidth]{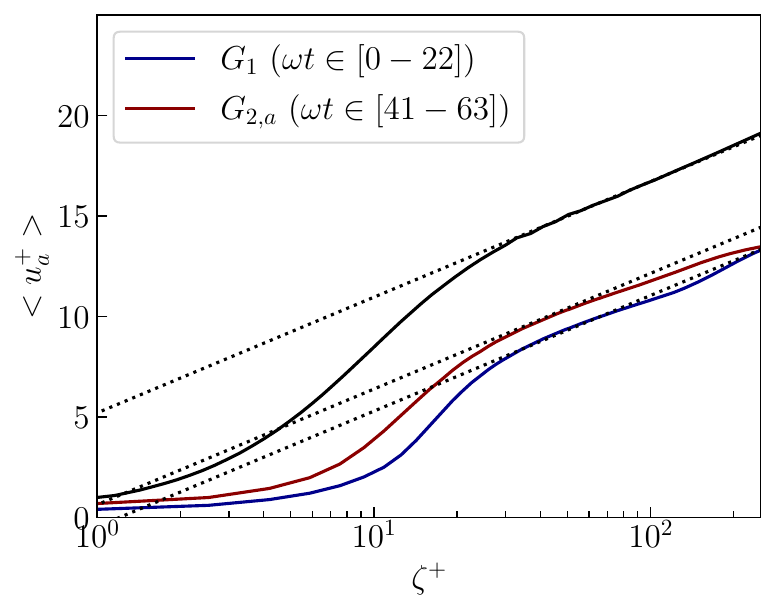}
  \put(-385,130){\small(a)}
  \put(-190,130){\small(b)}
  \caption{Streamwise velocity profile normalized by the friction velocity, $\langle u_a^+\rangle=\langle u_a\rangle/u_\ast $, as a function of vertical wave-following coordinate (in \revB{wall units}) $\zeta^+=\zeta u_\ast/\nu_a$ with $\nu_a=\mu_a/\rho_a$: (a) $u_\ast/c=0.5$ and (b) $u_\ast/c=0.9$, both at $Re_{\ast,\lambda}=214$ ($Re_\ast=720$). The velocity profiles are averaged over the same time windows as in figure~\ref{fig:profiles_0p5_0p9}. The dotted black lines refer to the fitted log law employed to estimate the intercept for each case. The continuous black line represents the mean velocity profile at $Re_\ast = 720$ for a flat stationary surface.}
  \label{fig:profiles_log_0p5_0p9}
\end{figure*}
The breaking-induced shift can be quantified considering the velocity profiles in semi-logarithmic form and rescaled in \revB{wall units}, as shown in Figure~\ref{fig:profiles_log_0p5_0p9}. 
The velocity profiles display a consistent up-shift which extends for the entire inner layer~\citep{Pope_2000,cimarelli2023wind}, composed of the viscous sublayer ($0<\zeta^+<5$), the buffer layer ($5<\zeta^+<30$) and log-law region ($\zeta^+>30$), where $\zeta^+$ is the vertical coordinate in \revB{wall units} $\zeta^+=\zeta u_\ast/\nu_a$ with $\nu_a=\mu_a/\rho_a$. The upshift is associated with the change in the instantaneous steepness due to breaking, leading to an overall drag reduction in the flow. Note that this upshift in the logarithmic region is mainly confined near the wave field ($30<\zeta^+\lesssim 100$). For larger $\zeta^+$, the mean velocity profile in the post-breaking stage, i.e. $G_{2,a}$, approaches the one in the pre-breaking stage, i.e. $G_1$, confirming that the airflow modulation induced by wave breaking is negligible in this region. \par 
It is also worth mentioning that the streamwise velocity profile at the same $Re_\ast$ for a flat stationary wall is well above the one with moving waves. Indeed, flows over waves experience much larger drag due to the added contribution of the pressure component, which is zero for flows over a flat wall~\citep{belcher1998turbulent}. This observation agrees with experimental works where a downshift of the velocity profile was observed as the wave steepness was increased~\citep{buckley2016structure,buckley2020surface}. \par
Using the velocity profile in logarithmic form, we can quantify this upshift induced by the wave breaking. For this purpose, we express $\langle u_a\rangle$ using the wall-normal coordinate, 
\begin{equation}\label{eqn:ua1_rough_noB}
  \dfrac{\langle u_{a,f}\rangle - \langle u_{a,i}\rangle}{u_\ast} = \dfrac{1}{\kappa}\log\left(\dfrac{\zeta_f^+-\zeta_i^+}{z_{0}^+}\right)\mathrm{,}
\end{equation}
where $\kappa$ is the Von Kármán constant, $z_{0}^+$ is the surface roughness, and $\zeta_{i,f}^+$ are the initial and final extents of the log-law region, i.e. $\zeta_i^+$ and $\zeta_f^+$. The corresponding velocities at these positions are termed $\langle u_{a,i}\rangle$ and $\langle u_{a,f}\rangle$, respectively. Using the velocity profile in the logarithmic region, both $\kappa$ and $z_0^+$ can be estimated using a least-square fit procedure~\citep{lin2008direct,sullivan2000simulation} applied to the growing and breaking regimes. To evaluate $z_{0}^+$ and $\kappa$, we consider $\zeta_i^+=30$ for all the stages, which corresponds, \revA{in physical units}, to $\zeta_i=0.56\lambda$, $0.28\lambda$ and $0.14\lambda$ for $Re_{\ast,\lambda}=53.5$, $107$ and $214$, respectively. The final extent $\zeta_f^+$ is taken equal to a common value of $\zeta_f^+\approx Re_{\ast,\lambda}/2$ to capture the breaking-induced drag reduction, which is confined in the near wave region, as clearly shown in fig.~\ref{fig:profiles_log_0p5_0p9}. Note that such $\zeta_f^+$ corresponds to a physical reference height of $\zeta_f=\lambda/2$, similar to the reference height employed to compute the wind-wave growth in~\citep{jeffreys1925formation,donelan2006wave} and to the height to compute the aerodynamic drag coefficient in \S~\ref{sec:drag}. \par
Results are reported in table~\ref{tab:roughness}, we note that the surface roughness slightly increases with $u_\ast/c$, whereas when the wave field breaks, its value drops and decreases with $u_\ast/c$. This trend is also confirmed during the second breaking cycle for the cases where it is available, although the change in $z_0^+$ during breaking is smaller due to the reduced breaking strength. Note that $z_0^+$ immediately after the breaking stage, i.e., over $G_{2,a}$, is smaller than $z_0^+$ evaluated before the second breaking, i.e. $G_{2,b}$ (when available), as the wave field has grown between the two windows. It is worth remarking that irrespective of the cases, we found a common value for the Von Kármán constant $\kappa\approx 0.40$, in agreement with the employed value in ocean and atmosphere models. \par
We note that the drag reduction mentioned here is studied at a fixed $Re_{\ast,\lambda}$ and with $(L_0-h_w)/\lambda=3.36$. To assess the sensitivity of the results on these parameters, we perform three additional simulations at $u_\ast/c=0.9$, two at $Re_{\ast,\lambda}=53.5-107$ with $(L_0-h_w)/\lambda=3.36$ and one at $Re_{\ast,\lambda}=107$ with $(L_0-h_w)/\lambda=6.72$. Results are discussed in the Appendix~\ref{app:Re_ast} and~\ref{app:num_wave}, and the surface roughness of these cases is reported in table~\ref{tab:roughness}. We observe a breaking-induced drag reduction of comparable magnitude to that measured at the largest $Re_{\ast,\lambda}$. This supports that the drag modulation induced by wave breaking is primarily controlled by $u_\ast/c$ and is less sensitive to $Re_{\ast,\lambda}$ and $(L_0-h_W)/\lambda$.
\begin{table}%[h!]
\centering
%\begin{tabular}{cccccccccc}
\begin{tabular}{ccccccc}
%\toprule
\toprule
{{$Re_{\ast,\lambda}$}} & {{$(L_0-h_W)/\lambda$}} & {{$u_\ast/c$}} & {{$z_{0,G_1}^+$}} & {{$z_{0,G_{2,a}}^+$}} &
{{$z_{0,G_{2,b}}^+$}} & {{$z_{0,F}^+$}} \\ %& {{$kz_{0,G_1}$}} & {{$kz_{0,G_{2,a}}$}} & {{$kz_{0,G_{2,a}}$}} & {{$kz_{0,F}$}} \\
\toprule
214  & 3.36 & 0.3 & 1.05 & -    & -    & -    \\ %& -                  & -                  & - & - \\
214  & 3.36 & 0.4 & 1.10 & 0.87 & -    & -    \\ %& 3.1$\cdot 10^{-2}$ & 2.8$\cdot 10^{-2}$ & 3.1$\cdot 10^{-2}$ & 2.8$\cdot 10^{-2}$ \\
214  & 3.36 & 0.5 & 1.14 & 0.83 & -    & -    \\ %& 3.4$\cdot 10^{-2}$ & 2.5$\cdot 10^{-2}$ & 3.4$\cdot 10^{-2}$ & 2.5$\cdot 10^{-2}$ \\
214  & 3.36 & 0.7 & 1.18 & 0.80 & 1.07 & 0.89 \\ %& 3.8$\cdot 10^{-2}$ & 2.3$\cdot 10^{-2}$ & 3.4$\cdot 10^{-2}$ & 2.5$\cdot 10^{-2}$ \\
214  & 3.36 & 0.9 & 1.21 & 0.77 & 1.15 & 0.85 \\ %& 4.0$\cdot 10^{-2}$ & 2.1$\cdot 10^{-2}$ & 4.0$\cdot 10^{-2}$ & 2.1$\cdot 10^{-2}$ \\
\midrule
53.5 & 3.36 & 0.9 & 0.26 & 0.17 & -    & -    \\ %& 3.0$\cdot 10^{-2}$ & 2.0$\cdot 10^{-2}$ & 3.0$\cdot 10^{-2}$ & 2.0$\cdot 10^{-2}$ \\
107  & 3.36 & 0.9 & 0.55 & 0.36 & 0.50 & 0.40 \\ %& 3.3$\cdot 10^{-2}$ & 2.2$\cdot 10^{-2}$ & 3.3$\cdot 10^{-2}$ & 2.2$\cdot 10^{-2}$ \\
107  & 6.72 & 0.9 & 0.57 & 0.38 & 0.52 & 0.42 \\ %& 3.3$\cdot 10^{-2}$ & 2.2$\cdot 10^{-2}$ & 3.3$\cdot 10^{-2}$ & 2.2$\cdot 10^{-2}$ \\
\bottomrule
%\bottomrule
\end{tabular}
\caption{Surface roughness for the stages $G_1$, $G_{2,a}$, $G_{2,b}$ and $F$, as defined in section~\ref{sec:Erms_evol}, see figure~\ref{fig:E_rms}(b). The surface roughness is expressed in viscous (or plus) units, $z_0^+=z_0u_\ast/\nu_a$ with $\nu_a=\mu_a/\rho_a$. The values are extracted from the velocity profiles in log form using a best-fit procedure reported in figure~\ref{fig:profiles_log_0p5_0p9}. For the case, $u_\ast/c=0.3$, only one surface roughness value is reported since the wave field, in this case, is in equilibrium with the flow. For the cases at $u_\ast/c=[0.7-0.9]$ at $Re_{\ast,\lambda}=214$ and for the case $u_\ast/c=0.9$ at $Re_{\ast,\lambda}=107$ with $(L_0-h_W)/\lambda=6.72$, $z_0^+$ is also reported for the second breaking cycle.}
\label{tab:roughness}
\end{table}
\section{Drag coefficients over growing and breaking waves}\label{sec:drag_coeff}
In this section, we connect our analysis of the momentum flux under growing and breaking waves at high wind speeds to formulations of the drag coefficient. We split our analysis between the growing waves and the breaking waves, for different $u_\ast/c$ and $Re_{\ast,\lambda}$. Two formulations are discussed: one based on first principles and directly integrating the momentum flux, and the one used in the laboratory and field observations.
\subsection{Estimation of the aerodynamic drag coefficient}\label{sec:drag}
In section~\ref{sec:mom_bud}, we showed that the pressure drag force $\tau_{p,x}$ increases in time as long as the wave field grows (associated with increased wave slope) and suddenly decreases when the wave field breaks. In section~\ref{sec:vel_prof}, we observed that the breaking event is associated with a flow acceleration, i.e. $\langle u_a\rangle/u_\ast$ shift towards larger values, and airflow separation occurs over the wave crests and troughs. To discuss the relative strength of the variation of the pressure force with the relative strength of mean flow variation, we introduce a dimensionless aerodynamic drag coefficient $C_{D,a}$, 
\begin{equation}\label{eqn:Cda}
  C_{D,a} = \dfrac{2\overline{\tau}_{p,x}}{\rho_a\overline{U}_{ref}^2(\overline{\zeta}=\zeta_{ref})A_\Gamma}\mathrm{.}
\end{equation}
where $\overline{\tau}_{p,x}$ is the time-averaged mean momentum flux associated with pressure,
\begin{equation}\label{eqn:tau_pm}
  \overline{\tau}_{p,x} = \dfrac{1}{\Delta T_{G,B}}\int_{\Delta T_{G,B}} \tau_{p,x} dt\mathrm{,}
\end{equation}
where $\Delta T_{G,B}$ is the time interval when the wave grows or breaks, respectively, and the integral is taken over this time interval. An equivalent definition to~\eqref{eqn:tau_pm} is employed for $\overline{U}_{ref}$. We split the dynamics into the growing and breaking stages to understand their respective contribution to the total aerodynamic drag, with the same convention as before, so that the length of the growth and breaking intervals varies with $u_\ast/c$. \par
The choice of the reference far-field boundary layer velocity $\overline{U}_{ref}$ evaluated at a reference height $\overline{\zeta}=\zeta_{ref}$ is based on two requirements: (i) being sufficiently far from the wave surface to always reside in the air during growth and breaking; (ii) being sufficiently close to the wave field to be affected by its dynamics (growth and breaking). It follows that $\overline{U}_{ref}$ increases as the flow accelerates in the breaking stage. Here, the requirements are satisfied typically if $\lambda/4 < \zeta_{ref} < \lambda$, and we consider $\zeta = \lambda/2$, similar to wind-wave growth sheltering discussion~\citep{jeffreys1925formation,donelan2006wave}. Different $\zeta_{ref}$ within the interval only changes the magnitude $C_{D,a}$, without altering the discussion. \par
\begin{figure*}%[t!]
  \centering
  \includegraphics[width=10.5 cm]{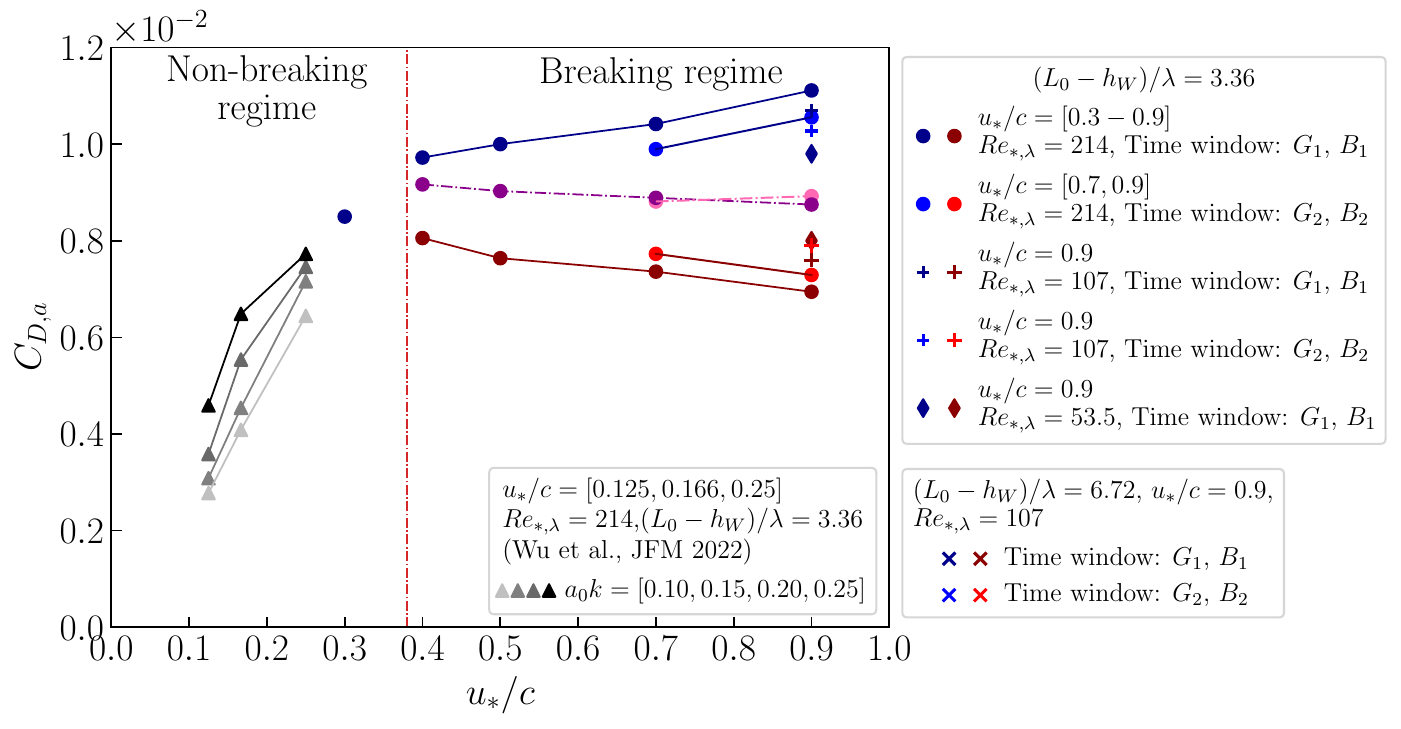}
  \caption{Aerodynamic drag coefficient $C_{D,a}$ (defined by eq.~\eqref{eqn:Cda}) for different $u_\ast/c$ in the growing (blue colors) and breaking (red colors) time intervals. For $u_\ast/c<0.35$, the simulated wave field is only growing (one would have to run the simulations longer to obtain breaking), so that all data are in the growing regime (and include increasing $a_0k$ see \citep{wu2022revisiting}). For $u_\ast/c>0.35$, both growing and breaking are present, and both ranges are separated by the red dashed-dot line. Whenever available, the data pertaining to the second growing and breaking cycles $G_2$, $B_2$ are displayed. The growing and breaking stages $G_1$, $G_2$, $B_1$ and $B_2$, are defined in figure~\ref{fig:E_rms}(b). Growing dynamics display a systematic increase in drag with $u_\ast/c$, while breaking induces a decrease in drag with increasing $u_\ast/c$. The average of the breaking and growing cycles are indicated in magenta, and we observe a saturation of the averaged drag at high wind speed.}
  \label{fig:CdA}
\end{figure*}
Figure~\ref{fig:CdA} shows the estimated $C_{D,a}$ as a function of $u_\ast/c$ and for both growing and breaking stages. We also re-analyze the non-breaking growing simulations from~\citet{wu2022revisiting} at lower $u_\ast/c$ and varying initial slope. In the non-breaking growing regime ($u_\ast/c < 0.35$), $C_{D,a}$ increases with $u_\ast/c$, with different values of $C_{D,a}$ at the same $u_\ast/c$ corresponding to different initial wave steepness (see also \citet{wu2022revisiting}). \par
For the range of $u_\ast/c$ that includes growth and breaking ($u_\ast/c > 0.35$), $C_{D,a}$ follows an increasing trend with $u_\ast/c$ in the growing (pre-breaking) stage (blue symbols) and a decreasing trend in the breaking stage (red symbols). Averaging $C_{D,a}$ over both the growing and breaking stages, the drag coefficient exhibits a saturation with a slight decrease at the highest wind speed (magenta symbols). Similar results are observed during the second breaking cycle, as shown in figure~\ref{fig:CdA} for the available cases ($u_\ast/c=[0.7-0.9]$ at $Re_{\ast,\lambda}=214$ and $u_\ast/c=0.9$ at $Re_{\ast,\lambda}=107$) with a slightly smaller drag coefficient, compared to the first breaking cycle. Thus, we can expect that averaging over multiple growing-breaking cycles would lead to a similar overall behavior. Finally, the results are not very sensitive to the friction Reynolds number $Re_{\ast,\lambda}$ and the number of waves in the simulations box $(L_0-h_W)/\lambda$. \par
%
%\revA{We remark that the present analysis is in physical agreement with the discussion in~\citep{sullivan2018turbulent,wu2022revisiting} on the control of the form drag by the wave slope, $a_\mathrm{rms}k$, for steep non-breaking waves. In particular, in this work, we showed that the wave slope saturates due to breaking, which then controls the saturation of the drag coefficient when presented as a function of $u_\ast/c$.} \par
%
From the analysis presented in this section, we can conclude that the aerodynamic drag saturation is driven by breaking events, which cause a marked reduction in the pressure drag, airflow separation, and a loss of coherence between the wave and the pressure fields. %\revA{Note that the present analysis can be related to the work in~\citep{sullivan2018turbulent,wu2022revisiting} on the control of the form drag by the wave slope, $a_\mathrm{rms}k$, for steep non-breaking waves. In the present work, we showed that the breaking event, which is responsible for wave slope saturation controls the saturation of the drag coefficient when presented as a function of $u_\ast/c$.}
%
% Obs CD
%
\subsection{Comparison of the drag coefficient to laboratory and field measurements}\label{sec:CD_f_obs}
The above discussion suggests that in laboratory and field measurements analysis, the drag saturation at high wind speed is the consequence of the spatial and temporal averages over breaking events and wave growth. In this section, we can qualitatively compare our results, obtained for narrow banded idealized conditions to the more complex multiscale observations at high wind speed~\citep{janssen2004interaction,donelan2004limiting,bye2006drag,sroka2021review,curcic2020revised,buckley2020surface}, by considering the classic oceanographic definition of the drag coefficient at $10$-m height, termed $C_D$. As mentioned in \S~\ref{sec:intro}, eq.~\ref{eqn:intro_cd}, $C_D$ is related to the definition of the total momentum flux $\tau_{t,x} = \rho_aC_D U_{10}^2$ with $\tau_{t,x}=\rho_au_\ast^2$ since at $10$-m height the airflow and the waves are assumed to be in equilibrium. For neutral atmospheric conditions, the velocity profile at $\overline{z}$ can be assumed to follow the logarithmic law of the wall \cite{edson2013exchange,ayet2022dynamical}, i.e. $U_{10} = (u_\ast/\kappa)\log \left(\overline{z}/z_0\right)$ as in equation~\eqref{eqn:ua1_rough_noB}, and $C_D$ becomes:
\begin{equation}\label{eqn:fin_CD}
  C_D(\overline{z}=10\,\ \mathrm{m}) = \dfrac{\kappa^2}{\log^2\left(\overline{z}/z_0\right)}\mathrm{.}
\end{equation}
Using the DNS data generated for this work and the one reported in~\cite{wu2022revisiting}, we compute $C_D$ from equation~\eqref{eqn:fin_CD} by using the velocity profiles evaluated during the first and a fraction of the second growing cycles for the different $u_\ast/c$ and $Re_{\ast,\lambda}$ (and in the second growing/breaking cycle when available). We convert the velocity profile in physical units by considering the length and timescale of the numerical system defined based on the air-water properties and the wave scales given by the Bond number. Using a best-fit procedure on the equation of the log profile ($\kappa=0.40$, as discussed in \S~\ref{sec:vel_prof}), a dimensional friction velocity and roughness $z_0$ can be retrieved, and $C_D$ can be evaluated using eq.~\eqref{eqn:fin_CD}. Note that we checked a-posteriori that this analysis conserves the ratio of the friction velocity over the wave phase speed. The calculation of $C_D$ is done on two averaging windows representative of the flow during the first growing cycle ($G_1$) and the disturbed flow just after breaking ($G_{2,a}$), since defining a logarithmic velocity profile requires a quasi-stationary profile. Such an approach differs slightly from the one employed for $C_{D,a}$, but we will see that the physical conclusions are identical, while the 10-m-based drag coefficient allows us to compare to existing laboratory data. 
\begin{figure*}%[t!]
  \centering
  \includegraphics[width=0.80\textwidth]{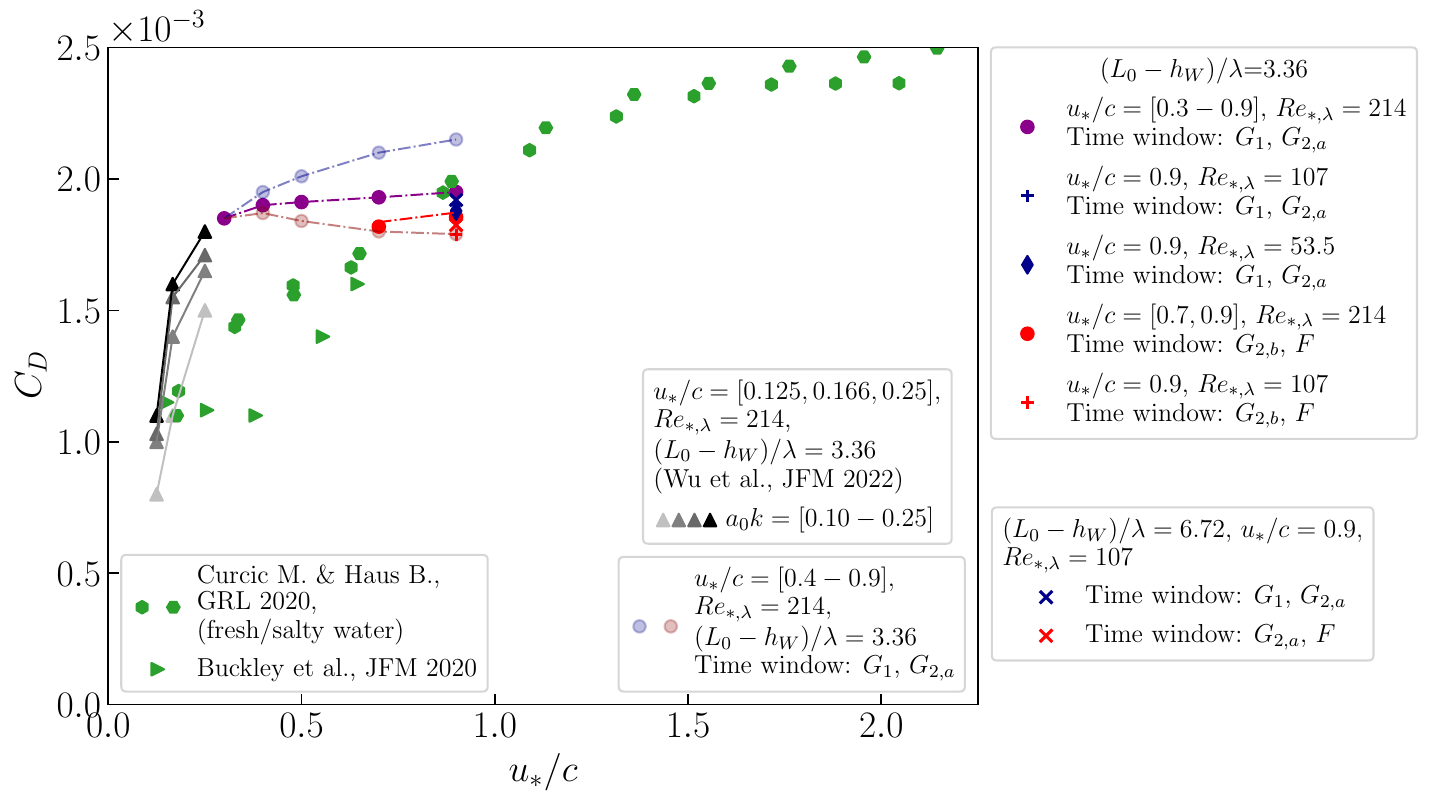}
  \includegraphics[width=0.80\textwidth]{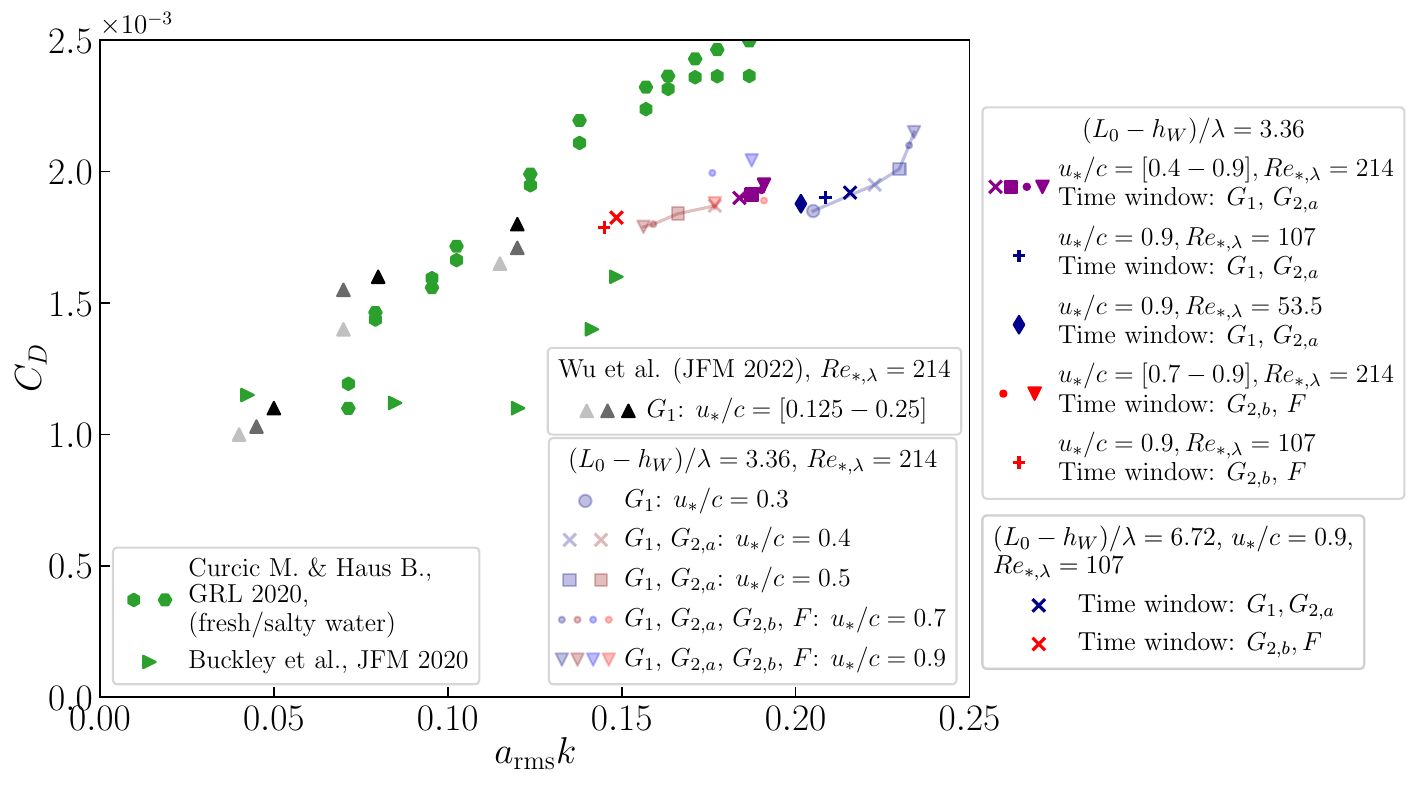}
  \caption{Drag coefficient $C_D$ evaluated at $\overline{z}=10$ $\mathrm{m}$ using equation~\eqref{eqn:fin_CD} as a function of $u_\ast/c$ (top) and $a_\mathrm{rms}k$ (bottom) with $a_\mathrm{rms}=a/\sqrt{2}$. The figure includes the calculation of $C_D$ with the data generated in this work and the one retrieved from~\citep{wu2022revisiting}. For all the cases, the reported $C_D$ is an average value between the first growing, $G_1$, and a fraction of the second growing cycle, $G_{2,a}$ (immediately after the breaking event). Whenever available, the data pertaining to the second growing cycle $G_{2,b}$ and the final stage $F$ are displayed. The employed time window to define $C_D$ follows the convention given in figure~\ref{fig:E_rms}(b). For the cases at $u_\ast/c=[0.4-0.5-0.7-0.9]$ at $Re_{\ast,\lambda}=214$ with $(L_0-h_W)/\lambda=3.36$, we separate between these two stages (blue and red dots). The green symbols display the experimental datasets from~\citep{buckley2020surface} up to $u_\ast/c\approx 0.71$ and from~\citep{curcic2020revised} up to $u_\ast/c\approx 2.25$.}
  \label{fig:CD_fields}
\end{figure*}
The values of $C_D$ for different $u_\ast/c$ exhibit a similar qualitative trend to that of $C_{D,a}$ with $u_\ast/c$. We observe a systematic increase of the drag in the growing stage (dark triangles and blue symbols) with $u_\ast/c$, while the drag in the breaking stage (red symbols) decreases with $u_\ast/c$. When averaging over both growing and breaking regimes, we observe a saturation at high wind speed ($u_\ast/c>0.4$). As for the aerodynamic drag coefficient $C_{D,a}$, the results are not very sensitive to $Re_{\ast,\lambda}$ and $(L_0-h_W)/\lambda$, as discussed in the Appendix~\ref{app:Re_ast} and Appendix~\ref{app:num_wave}. \par 
It is remarkable that the numerical values of $C_D$ in figure~\ref{fig:CD_fields}, with $C_D$ between 0.2 and 0.8$\cdot 10^{-3}$ for $u_\ast/c<0.3$ and $C_D$ between 1.5 and 2.0$\cdot 10^{-3}$ for $u_\ast/c>0.4$, closely match those reported in laboratory experiments~\citep{buckley2020surface,curcic2020revised} for similar values of $u_\ast/c$ (shown in green symbols). In laboratory experiments at high wind speeds, the drag coefficient is determined by long-time averaging over multiple growing and breaking cycles, leading to drag saturation. \par
%
%\begin{figure*}%[t!]
%  \centering
%  \includegraphics[width=0.80\textwidth]{figures/Cd_obs_armsk.pdf}
%  \caption{Drag coefficient $C_D$ evaluated at $\overline{z}=10$ $\mathrm{m}$ using equation~\eqref{eqn:fin_CD} as a function of $a_\mathrm{rms}k$ with $a_\mathrm{rms}=a/\sqrt{2}$. The figure includes the calculation of $C_D$ with the data generated in this work and the one retrieved from~\citep{wu2022revisiting}. For all the cases, the reported $C_D$ is an average value between the first growing, $G_1$, and a fraction of the second growing cycle, $G_{2,a}$ (immediately after the breaking event). Whenever available, the data pertaining to the second growing cycle $G_{2,b}$ and the final stage $F$ are displayed. The employed time window to define $C_D$ follows the convention given in figure~\ref{fig:E_rms}(b). For the cases at $u_\ast/c=[0.4-0.5-0.7-0.9]$ at $Re_{\ast,\lambda}=214$ with $(L_0-h_W)/\lambda=3.36$, we separate between these two stages (blue and red symbols). The green symbols display the experimental datasets from~\citep{buckley2020surface} and from~\citep{curcic2020revised}.}
%  \label{fig:CD_fields_rms}
%\end{figure*}
%
We note that the saturation of $C_D$ occurs at lower $u_\ast/c$ and $a_\mathrm{rms}k$ in our simulations than in laboratory experiments. This discrepancy may be attributed to differences in setup: our forced narrowbanded wave field contrasts with the laboratory’s multiscale finite fetch conditions at high wind speeds, where averaging spans multiple breaking-growing cycles, and the partitioning between regimes at the same $u_\ast/c$ may differ from our simulations. Similarly, directly comparing realistic $u_\ast/c$ values in field conditions is challenging due to the broad-banded nature of the wave field in the open ocean at high wind speed conditions. Additionally, the lower wave Reynolds number in our simulations affects the effective growth rate and time to reach breaking conditions and makes direct comparison in physical units, e.g. actual wind speed and wavelength, more challenging. Nevertheless, the ability of the simulations to capture drag saturation at high wind speed suggests that we are indeed approaching the correct asymptotic limit for high Reynolds numbers. \par
\revA{The present analysis can be related to the discussion from ~\citet{sullivan2018turbulent,wu2022revisiting} on the control of the form drag by the wave slope, $a_\mathrm{rms}k$, for steep non-breaking waves. In the present work, we showed that the breaking event, which is responsible for wave slope saturation controls the saturation of the drag coefficient when presented as a function of $u_\ast/c$. \par
Moreover, the drag coefficient $C_D$ is shown as a function of $a_\mathrm{rms}k$ in figure~\ref{fig:CD_fields}. During the growth phase of the wave field, $C_D$ increases with $a_\mathrm{rms}k$ up to approximately $a_\mathrm{rms}k \approx 0.15$ for non-breaking waves with finite steepness~\citep{wu2022revisiting}. As $a_\mathrm{rms}k$ reaches larger values, wave breaking initiates around $a_\mathrm{rms}k \approx [0.20-0.23]$. Focusing on the growing phase of the wave field, $C_D$ increases with $a_\mathrm{rms}k$ for the different $u_\ast/c$; however, as we approach the largest values, $a_\mathrm{rms}$ saturates, leading to a corresponding saturation in $C_D$. Conversely, during the breaking stage, $a_\mathrm{rms}k$ decreases as $u_\ast/c$ increases, resulting in a reduction in $C_D$. When considering the mean value of $C_D$ over a complete growing and breaking cycle, the variations in both $C_D$ and $a_\mathrm{rms}$ become significantly smaller, converging to similar values, so that all data points in the saturated drag coefficient also have saturated slope and appear clustered on the graph. This behavior highlights that wave breaking dynamics not only govern the saturation of $C_D$ but also control the saturation of the wave slope. This discussion presents similarities to~\cite{davis2023saturation} showing saturation of the wave mean square slope during tropical cyclones observed from drifting buoys, suggesting that simple parameterizations of the drag coefficient based on a wave slope metric could be promising.} \par

We suggest that experimental and field analysis of the drag coefficient reporting a statistical analysis of the partitioning between breaking and growing (temporally and/or spatially) could help better understand, quantify, and eventually parameterize drag saturation at high wind speed and constrain momentum flux models leveraging breaking occurrence statistics~\cite{kudryavtsev2014impact}. \par
Finally, it is very important to remark that while the saturation and reduction of $C_D$ has sometimes been attributed to the production of sea sprays~\citep{bye2006drag,veron2015ocean}, in the current setup, the production of droplets is negligible during the breaking stage of the wave field. Yet, we observe that the drag coefficient, whether defined as $C_{D,a}$ and $C_D$, displays a saturation at high $u_\ast/c$ when wave breaking is considered. This aspect shows that wave-breaking events, which cause the marked reduction of the pressure stress and the associated flow separation and determine the loss of coherence between the pressure and the wave fields, are sufficient to saturate the drag coefficient at high $u_\ast/c$.
%
% Conclusion
%
\section{Conclusions}\label{sec:concl}
We used state-of-the-art two-phase-fluid direct numerical simulations of wind-forced breaking waves to get new insights into the processes controlling momentum flux at high wind speed. We consider the ratio of wind friction velocity to wave phase speed, $u_\ast/c$, ranging from $0.3$ to $0.9$. The wave field is initialized with an amplitude below the breaking threshold, allowing the wind to be the primary driver of wave growth until the breaking conditions are reached and simulate the growing-breaking wave life cycle. We analyze the momentum flux, separating the growth and breaking stages, and treating them as two distinct sub-processes, and revisit drag coefficient analysis. \par
The momentum flux analysis underscores the dominance of the pressure force over the viscous force throughout all stages of wave evolution. When breaking occurs, the pressure force sharply decreases, and this reduction is compensated by a sudden acceleration of the flow field to conserve momentum. The remaining terms in the momentum budget, including a convective term accounting for the relative velocity between the airflow and waves, and a viscous force contribution, show smaller magnitudes and are less affected. The modulation of airflow caused by wave breaking results in an upshift in the streamwise velocity profile, corresponding to the drop in pressure force. \par
The relative strengths of these effects are quantified using the aerodynamic drag coefficient $C_{D,a}$, defined as the ratio of mean pressure force to mean velocity at a reference height within the wave-boundary layer ($z=\lambda/2$). During wave growth, $C_{D,a}$ increases with $u_\ast/c$, while during breaking, it decreases with $u_\ast/c$. When averaging over both growing and breaking regimes, we obtain a saturation of the drag coefficient $C_{D,a}$ at high enough $u_\ast/c$. A similar behavior is obtained if one computes from the DNS results the drag coefficient commonly employed in air-sea interaction, $C_D$, which is based on wind friction velocity $u_\ast$ and the reference velocity at 10-m height, $U_{10}$. The trend in the DNS closely mirrors field observations and laboratory experiments at high wind speeds for $C_D$, and the magnitude of the drag in both DNS and laboratory are in remarkable agreement. We observe a difference of the critical $u_\ast/c$ for which the drag starts to saturate, which might be attributed to the differences in setup, narrow or broad banded wave field, as well as averaging procedure. \par
Our simulations demonstrate that the drag coefficient, whether defined as $C_{D,a}$ or $C_D$, saturates when wave-breaking events are included. Notably, the production of droplets remains negligible due to the moderate $Bo$, indicating that breaking-induced dynamics - such as reduced pressure stress, loss of coherence between the airflow and wave field, and airflow separation - are sufficient to cause drag saturation. \par
The present framework can now be used to explore the associated energy fluxes going into the water column, including wind-wave growth, energy dissipation by breaking, and oceanic turbulence development, further quantifying the growing-breaking life cycle of ocean waves under strong wind forcing.
%
% Acknowledgments and Funding}
%
\subsection*{Acknowledgments and Funding}
This work is supported by the \textit{National Science Foundation} under grant 2318816 to LD (Physical Oceanography program), the \textit{NASA Wind Vector Science Team}, grant 80NSSC23K0983 to LD. NS acknowledges the support of the fellowship \textit{High Meadows Environmental Institute Postdoctoral Teaching Program}. Computations were performed using the \textit{Stellar} machine, granted by the \textit{Cooperative Institute for Earth System modeling} (CIMES), and managed by Princeton Research Computing. This includes the Princeton Institute for Computational Science and Engineering, the Office of Information Technology’s High-Performance Computing Center, and the Visualization Laboratory at Princeton University. 
%
%\appendix
%
\appendix
\revC{
\section{Procedure for converting Cartesian to Wave-Following Coordinates}\label{app:map_cart}
In this Appendix, we detail the procedure to transform a generic field defined in a Cartesian coordinate system into a wave-following one. This transformation involves three steps:
\begin{enumerate}
  \item The simulation outputs different instantaneous two-dimensional fields, hereinafter termed  $q=f(x,z)$, which has been already averaged along the spanswise $y$ direction. Next, this generic two-dimensional field $q$ is transformed from a Cartesian coordinate system into a wave-follower coordinate using a one-dimensional mapping function~\citep{sullivan2000simulation,wu2022revisiting}:
\begin{equation}\label{eqn:wave_fol}
  \begin{bmatrix}
    x \\
    %y \\
    z
  \end{bmatrix}
    = \begin{bmatrix}
    x(\xi,\zeta) \\
    %y(\eta) \\
    z(\xi,\zeta)
  \end{bmatrix}
    = \begin{bmatrix}
    \xi \\
    %\eta \\
    \zeta + \overline{\eta}(\xi)\exp(-k|\zeta|)
  \end{bmatrix}\mathrm{,}
\end{equation}
where $\overline{\eta}(\xi)$ is the spanwise-averaged surface elevation function. For a sufficiently large value of $k|\zeta|$, $\exp(-k|\zeta|)\approx 0$ and the vertical Cartesian coordinate coincides with the vertical wave-following coordinate. Note that alternative mapping functions to~\eqref{eqn:wave_fol} have been proposed, but the sensitivity of the results on the choice of mapping is typically negligible~\citep{sullivan2000simulation};
  \item Each wave-following two-dimensional field is averaged along the periodic direction $x$, including also the regions near the wave troughs; 
  \item The obtained vertical profile, i.e. $q=g(\zeta)$, is time-averaged over the cycles $G_1$ and $G_{2,a}$ as defined in figure~\ref{fig:E_rms}(b). Note that during $G_1$ and $G_{2,a}$ ($G_{2,b}$ and $F$), the airflow is quasi-stationary, and employing time-averaging is an adequate approach to define turbulence statistics.
\end{enumerate} 
}
\section{Calculation of the momentum fluxes}\label{app:fluxes}
We provide details on how we evaluate the momentum fluxes in equation~\eqref{eqn:u_a}, i.e. $\rho_a\phi_{c,x}$, $\tau_{p,x}$ and $\tau_{\nu,x}$. These fluxes at the boundaries are not zero only at the two-phase interface, since at the other faces of the computational box (shown in figure~\ref{tab:set_up}), they are zero due to periodicity along the horizontal directions and at the top boundary due to the imposed no-penetration/free-slip condition. \par
The calculation of the momentum fluxes is conducted in two steps. First, the interface $\Gamma$ is slightly shifted vertically by $\Delta_s=0.1/k$, corresponding approximately to $4$ grid points at a resolution $\mathrm{Le}=10$. As noted by~\citet{wu2022revisiting}, this step is a numerical strategy to ensure that the pressure and the velocity gradients are evaluated in the air domain only. As long as $\Delta_s$ stays within the viscous sub-layer, its numerical value has a negligible impact on the evaluation of the different terms of the momentum budget. Next, the surface integrals in equations~\eqref{eqn:Ftx} and~\eqref{eqn:u_a} are transformed into volume integrals using the Gauss theorem, e.g. $\int_\Gamma p\mathbf{n}\cdot\mathbf{e}_x dS=\int_{\Omega_a}\nabla\cdot(p\mathbf{e}_x) dV=\int_{\Omega_a}(\partial p/\partial x) dV$. This surface-to-volume integral transformation allows evaluating the momentum fluxes over the air volume $\Omega_a$ rather than on $\Gamma$. Owing to the VoF reconstruction, $\Omega_a$ remains a conserved quantity in our simulations, even when the wave field breaks, whereas $\Gamma$ does not. Importantly, the volume and surface integral formulations yield the same numerical results when the wave field grows. 
\section{Convergence studies between $\mathrm{Le}=10$ and $\mathrm{Le}=11$}\label{app:conv}
We perform a grid convergence study for wave energy and momentum flux for two cases at $u_\ast/c=0.9$ and $Re_{\ast,\lambda}=214$ with $(L_0-h_W)/\lambda=3.36$ and $(L_0-h_W)/\lambda=6.72$ ($4$ and $8$ waves per box size, respectively). Two levels of refinement are considered, $\mathrm{Le}=10$ and $\mathrm{Le}=11$. At $\mathrm{Le=10}$, we have $256$ ($128$) and $512$ ($256$) grid points per wavelength in the case of $(L_0-h_W)/\lambda=3.36$ ($(L_0-h_W)/\lambda=6.72$) (and twice as much at Le=11). As we will see, the number of grid points per wave boundary layer (as defined in~\citet{mostert2022high}) is enough to obtain grid convergence of energy dissipation due to breaking. \par
\begin{figure*}%[t!]
  \centering
  \includegraphics[width=0.45\textwidth]{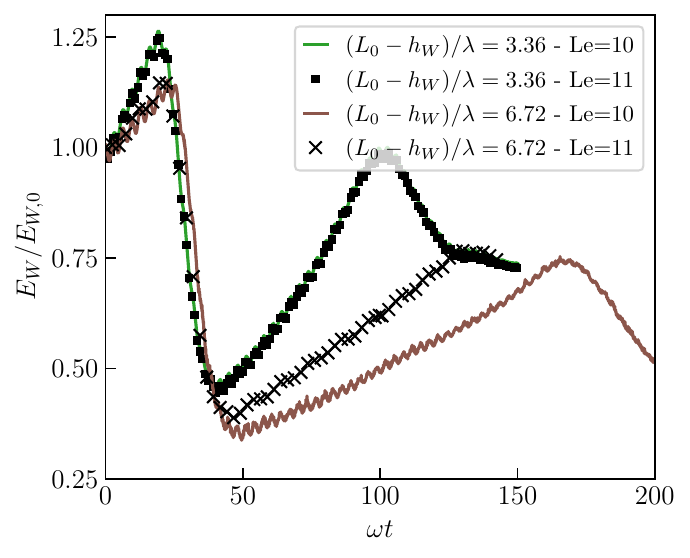}
  \caption{Evolution of the normalized wave energy $E_W/E_{W,0}$ as a function of the dimensionless time $\omega t$ with $E_{W,0}=E_W(t=0)$ for $Re_{\ast,\lambda}=107-214$ at $u_\ast/c=0.9$. Note that these two cases share the same $Re_\ast=720$ but different $(L_0-h_W)/\lambda=3.36-6.72$. The continuous color lines refer to $\mathrm{Le}=10$, the black symbols to $\mathrm{Le}=11$.}
  \label{fig:E_rms_L10_L11}
\end{figure*}
\begin{figure*}%[t!]
  \centering
  \includegraphics[width=0.8\textwidth]{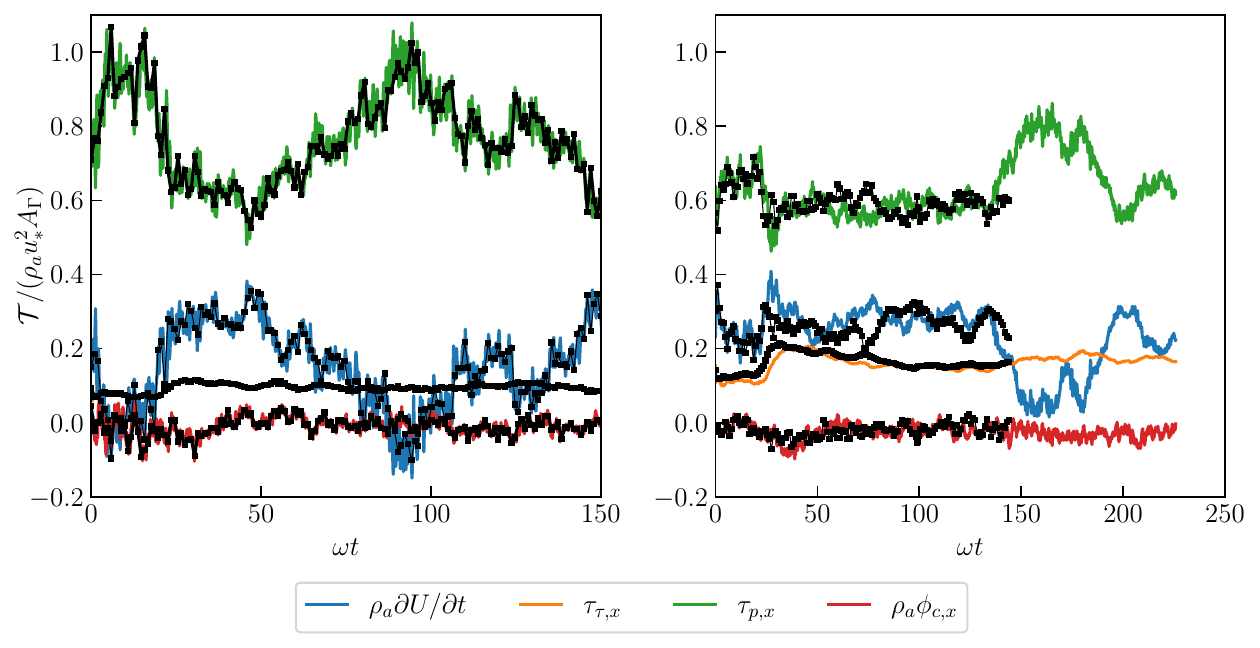}
  \put(-305,150){\small(a)}
  \put(-155,150){\small(b)} 
  \caption{Grid-convergence study for the momentum budget in the streamwise direction, as in equation~\eqref{eqn:u_a} for (a) $(L_0-h_W)/\lambda=3.36$ and $Re_{\ast,\lambda}=214$ and (b) $(L_0-h_W)/\lambda=6.72$ and $Re_{\ast,\lambda}=107$ at $u_\ast/c=0.9$. Both cases share the same $Re_\ast=720$. The colored curves refer to $\mathrm{Le}=10$; the symbols refer to resolution $\mathrm{Le}=11$. On the $y$-label $\mathcal{T}$ represents the variation in the instantaneous flow $\rho_a\partial U/\partial t$, the viscous stress $\tau_{\nu,x}$, the pressure stress $\tau_{p,x}$, the convective term $\rho_a\phi_{c,x}$ or the driving force $\Pi_f$. Each budget component is normalized by the total stress $\rho_au_\ast^2 A_\Gamma$.}
  \label{fig:mom_flux_L10_L11}
\end{figure*}

Figure~\ref{fig:E_rms_L10_L11} displays the normalized wave energy $E_W/E_{W,0}$ as a function of the dimensionless time $\omega t$. For $(L_0-h_W)/\lambda=3.36$, employing a maximum refinement level of $\mathrm{Le}=10$ guarantees grid-converged results across all wave field's growing and breaking stages (results are identical at Le=11). Moreover, this level of refinement proves adequate to achieve grid-independent results also for the momentum budget components, as shown in figure \ref{fig:mom_flux_L10_L11}(a).
Increasing $(L_0-h_W)/\lambda$ from $3.36$ to $6.72$, while keeping the other dimensionless parameters fixed, makes the grid requirement more demanding, as grid convergence still requires the same number of grid points per wavelength (or per wave boundary layer, which scales with the wavelength). In particular, during the first growing stage, the wave energy and the components of the momentum budgets are well captured at a resolution of $\mathrm{Le}=10$, as shown in figure~\ref{fig:E_rms_L10_L11} and figure~\ref{fig:mom_flux_L10_L11}(b). Conversely, a deviation with respect to $\mathrm{Le}=11$ occurs during the breaking stage and the second growing stages of the wave field. Accordingly, fixing all the other parameters, cases with $(L_0-h_W)/\lambda>6.72$ requires $\mathrm{Le}=11$ to obtain grid-independent results. \par
\section{Sensitivity of the momentum flux to $Re_{\ast,\lambda}$}\label{app:Re_ast}
We now present the results of a sensitivity study to the turbulent air-side friction Reynolds number, $Re_{\ast,\lambda}$. We consider three cases at a fixed $u_\ast/c=0.9$: $Re_{\ast,\lambda}=53.5$ and $Re_{\ast,\lambda}=107$ with $(L_0-h_W)/\lambda=3.36$, and $Re_{\ast,\lambda}=107$ with $(L_0-h_W)/\lambda=6.72$, which correspond to a friction Reynolds number $Re_\ast=180-360-720$, respectively. \par
\begin{figure*}%[t!]
  \centering
  \includegraphics[width=0.48\textwidth]{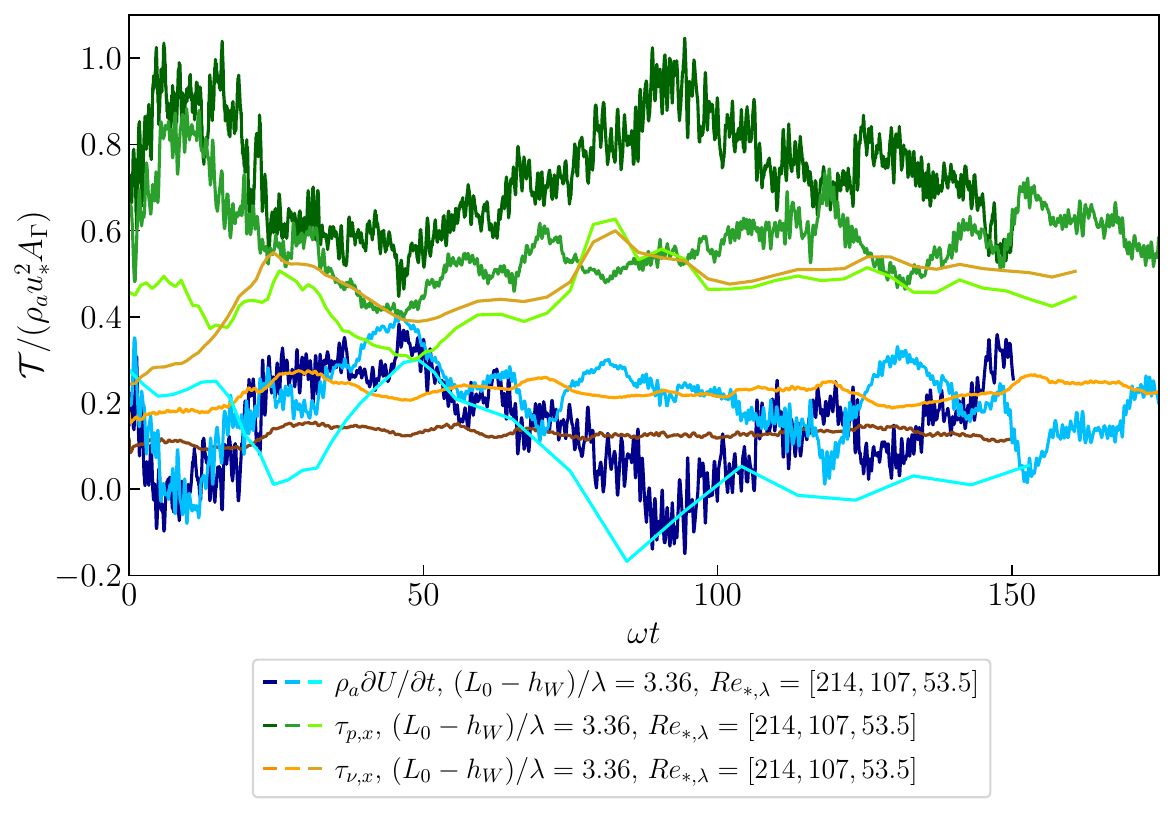}
  \includegraphics[width=0.48\textwidth]{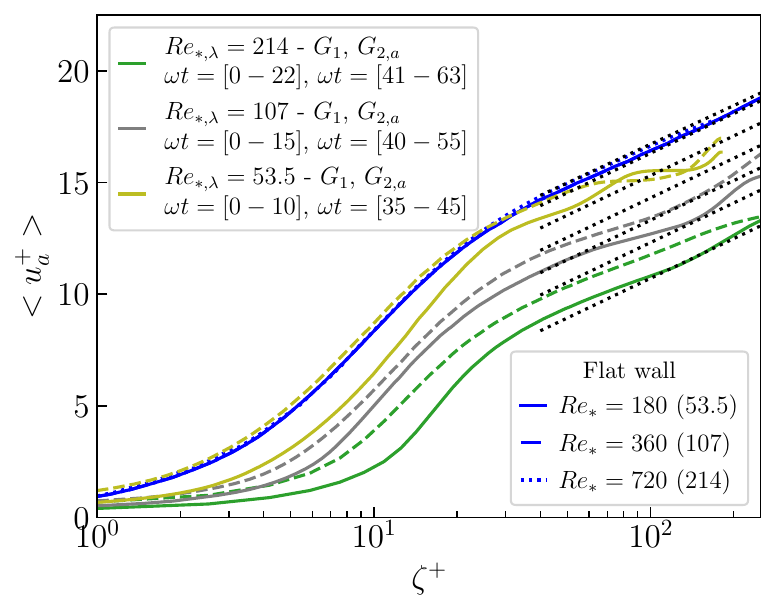}
  \put(-375,120){\small(a)}
  \put(-185,120){\small(b)} 
  \caption{(a) Contributions of the momentum budget in the streamwise direction, as in equation~\eqref{eqn:u_a}, for $Re_{\ast,\lambda}=53.5-107-214$ with $(L_0-h_W)/\lambda=3.36$. On the $y$-label of both panels, $\mathcal{T}$ represents the variation in the instantaneous flow $\rho_a\partial U/\partial t$, the viscous stress $\tau_{\nu,x}$ and the pressure stress $\tau_{p,x}$. Each budget component is normalized by the total stress $\rho_au_\ast^2A_\Gamma$. (b) Streamwise velocity profile normalized by the nominal friction, $\langle u_a^+\rangle=\langle u_a\rangle/u_\ast $, as a function of vertical wave-following coordinate (in \revB{wall units}) $\zeta^+=\zeta u_\ast/\nu_a$ for $Re_{\ast,\lambda}=53.5-107-214$ with $(L_0-h_W)/\lambda=3.36$ in colored lines (continuous and dashed lines for the first and second growing cycle, $G_1$ and $G_{2,a}$, respectively). The dotted black lines are the linear fit. In blue colour, we report the corresponding cases pertaining to the stationary and flat wall at the same $Re_\ast$ (the equivalent $Re_{\ast,\lambda}$ is displayed in the parenthesis). Note that the velocity profiles are almost indistinguishable in the case of a flat stationary wall at $Re_\ast=360-720$, while the one at $Re_\ast=180$ (blue dotted lines) is slightly upshifted.}
  \label{fig:diff_retau}
\end{figure*}
We start comparing the three cases by inspecting the momentum budget in the streamwise direction, as displayed in figure~\ref{fig:diff_retau}(a). The budget clearly shows that the time-varying pressure flux $\tau_{p,x}$ has a smaller magnitude over the entire growing and breaking cycles as $Re_{\ast,\lambda}$ decreases. A larger viscous flux, $\tau_{\nu,x}$ compensates for this decrease. \par
The overall reduced pressure drag experienced by the wave field at lower $Re_{\ast,\lambda}$ modifies the streamwise velocity profile, as displayed in figure~\ref{fig:diff_retau}(b). As we increase $Re_{\ast,\lambda}$, the velocity profile has a consistent downshift compared to the flat and stationary wall scenario (blue lines in figure~\ref{fig:diff_retau}(b), which is caused by the increased drag due to the pressure component (absent for a flat wall). When the wave breaks, in all the cases, there is an upshift of the velocity profile, and the associated drag reduction is solely induced by wave breaking. As visible in the figure, the upshift is largest for $Re_{\ast,\lambda}=214$, and it slightly decreases for $Re_{\ast,\lambda}=53.5-107$. Despite an unavoidable $Re_{\ast,\lambda}$-effect, this analysis confirms that the breaking-induced flow acceleration and consequent drag reduction is associated with the change in the wave steepness of the wave field and surface roughness and is not significantly affected by $Re_{\ast,\lambda}$.
\section{Sensitivity of the momentum flux on $(L_0-h_W)/\lambda$}\label{app:num_wave}
We present a sensitivity study on the ratio of the box size to the wavelength, i.e. $(L_0-h_W)/\lambda$, or equivalently, the number of waves per box size. To this end, we run an additional case at $Re_{\ast,\lambda}=107$ with $(L_0-h_W)/\lambda=6.72$ and $u_\ast/c=0.9$. Given the larger $(L_0-h_W)/\lambda=6.72$, we set $\mathrm{Le}=11$, as discussed in Appendix~\ref{app:conv}. 
\begin{figure*}%[t!]
  \centering
  \includegraphics[width=0.6\textwidth]{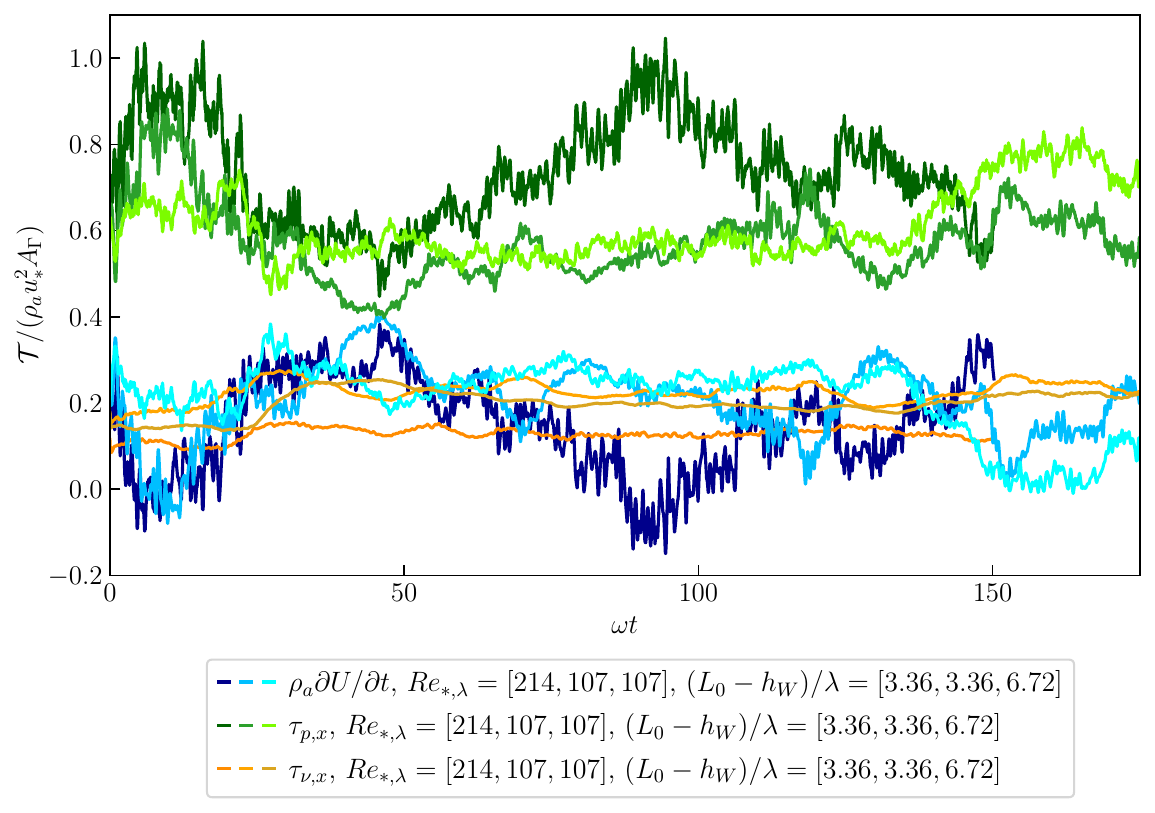}
  \caption{Contributions of the momentum budget in the streamwise direction, as in equation~\eqref{eqn:u_a}, for the following cases: $Re_{\ast,\lambda}=214$ with $(L_0-h_W)/\lambda=3.36$, $Re_{\ast,\lambda}=107$ with $(L_0-h_W)/\lambda=6.72$ and the case with $Re_{\ast,\lambda}=107$ with $(L_0-h_W)/\lambda=6.72$. On the $y$-label $\mathcal{T}$ represents the variation in the instantaneous flow $\rho_a\partial U/\partial t$, the viscous stress $\tau_{\nu,x}$ and the pressure stress $\tau_{p,x}$. The convective term $\rho_a\phi_{c,x}$ or the driving force $\Pi_f$ are omitted for clarity. Each budget component is normalized by the total stress $\rho_au_\ast^2A_\Gamma$.}
  \label{fig:mom_flux_k4_k8}
\end{figure*}
This additional case is compared with: (i) the case at different with $(L_0-h_W)/\lambda=3.36$ but equal $Re_{\ast,\lambda}=107$ and (ii) the case at different $Re_{\ast,\lambda}=214$ and $(L_0-h_W)/\lambda=6.72$, but equal $Re_\ast=720$. \par
Figure~\ref{fig:mom_flux_k4_k8} displays the different components of the momentum flux as in equation~\eqref{eqn:u_a}. We can see that despite the different values of $(L_0-h_W)/\lambda$, the cases with the same $Re_{\ast,\lambda}=107$ display very similar momentum fluxes. Conversely, the case at $Re_{\ast,\lambda}=214$ shows larger pressure and smaller viscous fluxes. These observations suggest two conclusions: first, while the ratio $(L_0-h_W)/\lambda$ imposes stricter resolution criteria, its effect on the results is limited. Second, when comparing momentum fluxes across different cases, the relevant dimensionless parameter is $Re_{\ast,\lambda}$. \par
\begin{figure*}%[t!]
  \centering
  \includegraphics[width=0.49\textwidth]{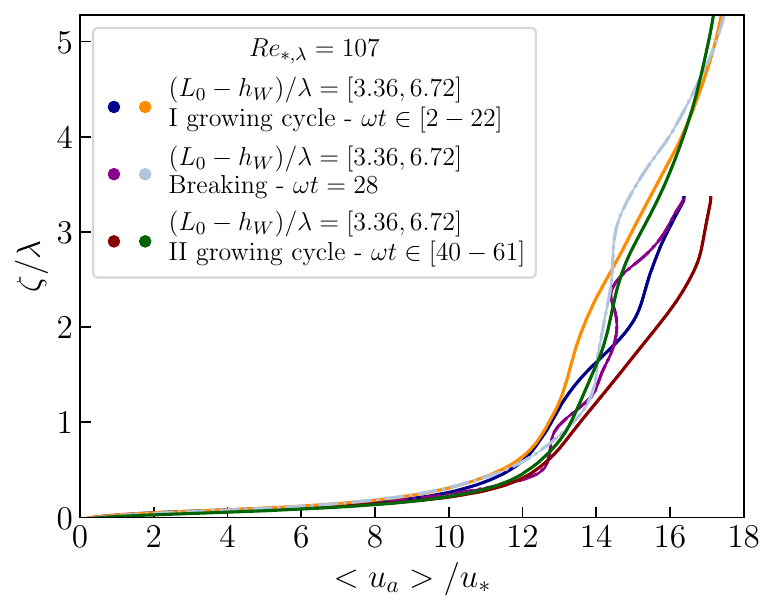}
  \includegraphics[width=0.49\textwidth]{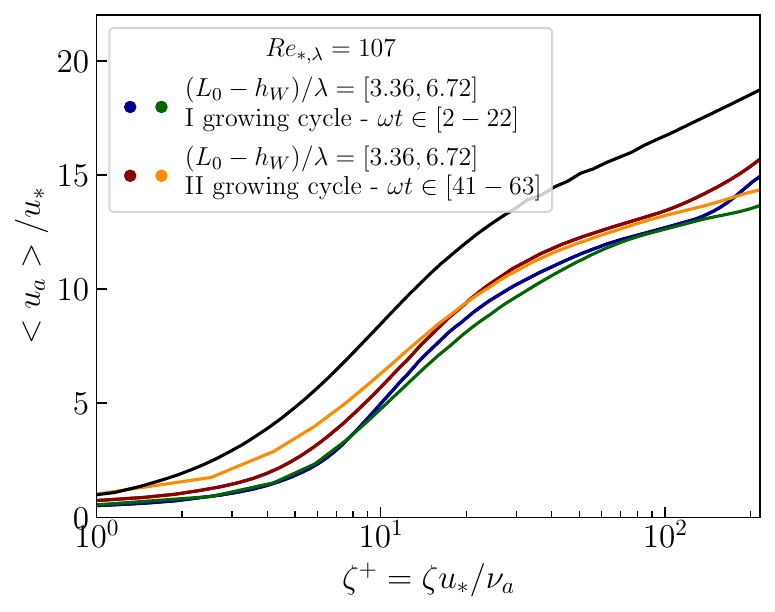}
  \put(-380,135){\small(a)}
  \put(-190,135){\small(b)}
  \caption{Streamwise velocity profile normalized by the nominal friction, $\langle u_a\rangle/u_\ast$ in (a) physical or (b) \revB{wall units} for the cases at $Re_{\ast,\lambda}\approx 107$ and $(L_0-h_W)/\lambda=3.36-6.72$. The solid black line in the panel (b) represents the velocity profile of the flow over a solid wall at the same $Re_{\ast,\lambda}=107$. Note that differently from the profiles in the panel (a), the velocity profiles as a function of the viscous unit do not collapse since the normalization factor for the viscous unit, i.e. $\nu_a/u_\ast$, is independent of the number of waves per unit of the box size.}
  \label{fig:log_profile_0p9_k4_k8}
\end{figure*}

Figure~\ref{fig:log_profile_0p9_k4_k8} reports the velocity profiles for the two cases at $Re_{\ast,\lambda}\approx 107$ with $(L_0-h_W)/\lambda=3.36-6.72$. The velocity profiles in figure~\ref{fig:log_profile_0p9_k4_k8}(a) of the two cases display a good collapse during the two considered growing cycles in the region immediate to the wave field up to $z/\lambda \approx 1$. Here, the profiles are similar since the two cases share the same $Re_{\ast,\lambda}$, which is the controlling parameter for the turbulent intensity near the wave field. Away from the wave field region, i.e. $z/\lambda > 1$, the velocity profile pertaining to the same growing cycle diverges and the cases with $(L_0-h_W)/\lambda=6.72$ is downshifted. Indeed, in the airflow regions well above the wave field, the turbulent intensity is controlled by $Re_\ast$, which is larger for the case at larger $(L_0-h_W)/\lambda$. \par
Figure~\ref{fig:log_profile_0p9_k4_k8}(b) reports the velocity profiles in semi-logarithmic form, which both display an upshift induced by wave breaking. The upshift is of similar magnitude for both cases, which further confirms that the breaking-induced drag reduction is mainly controlled by $u_\ast/c$, whereas $(L_0-h_W)/\lambda$ has a limited effect. 
\bibliographystyle{jfm}
\bibliography{bibfile.bib}
\end{document}